\documentclass[11pt,a4paper]{article}

\usepackage{amsmath,amssymb}
\usepackage{epsfig,graphicx}
\usepackage{graphicx}
\usepackage{rotating}
\usepackage{cancel}
\usepackage{bm}
\usepackage{color}
\usepackage{comment}
\usepackage{cite}
\usepackage{psfrag}
\usepackage{slashed}
\usepackage{soul}
\usepackage{tikz}
\usepackage{hyperref}
\usepackage{bbold}
\usepackage{siunitx}
\usepackage[normalem]{ulem}
\usepackage{subcaption}
\usepackage{adjustbox}
\usepackage{braket}

\renewcommand\[{\left[}

\newcommand{\exclude}[1]{}

\def\beq{\begin{equation}}
\def\eeq{\end{equation}}

\begin{document}
\title{\Large{\textbf{Constraining Multiple Kinetically Mixed Dark Photons}}}
\author{ Joerg Jaeckel$^{a}$ and Sebastian Monath$^{a,b,c}$\\[2ex]
$^{a}$\small{\em Institut f\"ur theoretische Physik, Universit\"at Heidelberg,} \\
\small{\em Philosophenweg 16, 69120 Heidelberg, Germany}\\
$^{b}$\small{\em Ecole Normale Superieure (ENS) de Lyon} \\
\small{\em 15 parvis René Descartes - BP 7000, 69342 Lyon Cedex 07, France}\\
$^{c}$\small{\em Université Claude Bernard Lyon I, Institut des 2 Infinis} \\
\small{\em 4 Rue Enrico Fermi, F-69622 Villeurbanne Cedex, France}\\
\vspace*{1.5cm}}
\date{}
\maketitle
\begin{abstract} 
\noindent 
Extra U(1) gauge bosons under which Standard Model particles are uncharged, aka dark photons, are a simple and well-motivated extension of the Standard Model. 
There could be a single, but also several or even many such dark photons.
However, most studies consider only a single dark photon. Here, we want to look at the more general case of multiple dark photons interacting with the Standard Model via kinetic mixing.
We consider a range of standard probes, Cavendish experiments, light-shining-through-walls experiments, as well as energy loss in stars. To explore the rather high-dimensional parameter space of the kinetic mixing matrix, we pursue a statistical approach, considering different distributions for the kinetic mixing parameters.
\end{abstract} 
\vspace{.3cm}
\noindent

\newpage
\section{Introduction}

The quest to discover physics beyond the Standard Model (SM) remains one of the central challenges for fundamental particle physics.
One simple, yet very plausible extension is to have an extra U(1) gauge factor under which the SM particles are uncharged. The new particle is then automatically only feebly interacting, hence a dark photon (DP). The leading interaction of the DP with the SM is via kinetic mixing with the hypercharge boson or the photon~\cite{Okun,Holdom,Foot:1991kb}, depending on the relevant energy scale. It therefore provides a portal to the hidden or dark sector (cf., e.g.~\cite{Binoth:1996au,Schabinger:2005ei,Patt:2006fw,Ahlers:2008qc,Batell:2009yf,Beacham:2019nyx,Agrawal:2021dbo,Antel:2023hkf}, for a discussion of this and other portals).
Unlike regular photons, they are not required to be massless. Instead, they may acquire a mass through the Higgs or Stueckelberg mechanism \cite{Englert:1964et,higgs1964broken,Stueckelberg:1938zz}. Indeed, in the absence of extra matter particles or higher-dimensional interactions~\cite{Dobrescu:2004wz,Dobrescu:2006au,Fabbrichesi:2020wbt,Coudarchet:2025dfd}, they are only observable if massive.

Dark photons are the objects of vivid experimental research and astrophysical/cosmological considerations, e.g.~\cite{cavendish,Loegl,boundary1,boundary2,boundary3,LSWAlps,Popov,Popov2,redondo08,redondo13,Pradler, redondo2009massive}, cf. also the reviews~\cite{Beacham:2019nyx,Agrawal:2021dbo,Antel:2023hkf}. However, when the outcomes of these experiments or astrophysical observations are translated into bounds of the dark photon mass and mixing parameters, one usually employs the assumption of a single dark photon model. While this simplification eases calculations significantly and admits many important conclusions, it is certainly not the most general setting.

Indeed, it is straightforward to imagine that there are multiple U(1) gauge factors (cf.~\cite{Masso:2006gc,Heeck:2011md} for early examples) or, as motivated by theories with extra dimensions, towers of Kaluza-Klein modes~\cite{Dienes,McDonald1,McDonald2,KaluzaWJR,Rizzo }. 
This may have important consequences for the phenomenology. For example, when considering the energy loss in stars, it may now be possible to excite a whole range of dark photons~\cite{KaluzaWJR}. But, as we will see, effects can also be more subtle. In laboratory experiments, such as light-shining-through-walls, that are based on the coherence of different modes, destructive interference can occur (see also~\cite{deGiorgi:2025ldc} for the case of axions). Consequently, regions of model space may be reopened to consideration, which were previously thought to be ruled out. 
The aim of this note\footnote{Part of this note was work done in the context of the Bachelor thesis of one of the authors (S. Monath).} is to generalize results, notably the experimental and observational constraints from the single DP setting, to arbitrary numbers of additional vector fields. In particular, we consider constraints from Cavendish experiments~\cite{cavendish,Loegl,boundary1,boundary2,boundary3}, light-shining-through-walls (LSW) experiments~\cite{Okun,Anselm:1985obz,VanBibber:1987rq,Anselm:1987vj,Popov,ALPSII:2025eri,Jaeckel:2010ni,LSWRedondo,LSW,Graham:2015ouw,Irastorza:2018dyq}, and solar energy loss~\cite{Popov,Popov2,redondo08,Pradler,redondo13,redondo2009massive} (see~\cite{Raffelt:1996wa} for \emph{the} review). While there exists a plethora of other experimental probes, we consider the present selection as illustrative of the effects of considering multiple dark photons and leave a more exhaustive investigation to future work.

An important challenge in delineating the phenomenology of many dark photons is that, with increasing number of DPs, the number of free parameters grows rapidly. In fact, for $n$ dark photons, the number of kinetic mixing parameters grows as $n(n+1)/2$. Additionally, we have $n$ mass parameters. This quickly becomes unwieldy.   
Therefore, we resort to a statistical treatment of the parameters. In practice, we make assumptions on the overall scales of the mixings and masses and then take the remaining parameters to follow different statistical distributions. We can then draw statements, e.g., on the fraction of excluded “models" for a given overall size of the kinetic mixing and the distribution around their median. 
This approach is closely related to the concept known in the literature as \textit{anarchy}, an approach originally used in neutrino physics \cite{Hall_2000}, which has recently been applied to the phenomenology of an arbitrary number of ALPs~\cite{Chadha-Day1, Chadha-Day2}\footnote{For a recent study of non-anarchical, multiple ALPs in LSW experiments see~\cite{deGiorgi:2025ldc}.} by using the Haar-measure to describe the ALPs' parameter distributions\footnote{These “axiverse" models arise naturally from string compactifications \cite{axiverse} and can also be constructed in a Kaluza-Klein setup \cite{Dienes2}. }. 
The statistical results from this approach are different from those 
derived for the Kaluza-Klein models in \cite{KaluzaWJR,McDonald1,McDonald2,Rizzo, Dienes}. This is because we do not prescribe any sequential relation for the masses and couplings of the dark Photons. Moreover, we also consider interference effects whenever the dark photons appear in an intermediate state, notably in light-shining-through-walls experiments.

The game plan for this note is as follows. In section~\ref{sec::2}, we present the generalized equations relating dark photon parameters to the physical observables measured in a selection of the most relevant experiments constraining dark photons. In particular, we point out when the effects of multiple DPs are additive and when there is potential for destructive interference. In section~\ref{sec::3}, we perform a statistical analysis considering two cases: 1) assuming that the DPs do not kinetically mix among themselves; 2) permitting kinetic mixings between the individual new fields but also requiring that there are no unphysical instabilities such as ghosts. We also consider how the phenomenology depends on the number of DPs.
Ultimately, we briefly recapitulate and discuss our results in section~\ref{sec::disc}.

\section{Calculating observables for multiple dark photons} \label{sec::2}

The effective Lagrangian describing the interaction between classical QED and several hypothetical dark photons with masses $m_i$, couplings $\epsilon_i$ and mixings $\chi_{ij}$ is given by,
\begin{equation}\label{eq::1}
        \mathcal{L} = -
\frac{1}{4} F_{\mu\nu}F^{\mu\nu}-\frac{1}{4}{X_i}_{\mu\nu}X_i^{\mu\nu}-\frac{1}{4}\chi_{ij}{X_i}_{\mu\nu}X_i^{\mu\nu} -\frac{\epsilon_i}{2}F_{\mu\nu}X_i^{\mu\nu}+\frac{m_{X_i}^2}{2}{X_i}_{\mu}X_i^{\mu}+j_{\mu}A^{\mu}-\frac{1}{2}A_\mu\Pi^{\mu\nu}A_{\nu}
\,.
\end{equation}
This is a well-established (cf., e.g.,~\cite{Masso:2006gc}) and straightforward generalization of kinetic mixing for a single DP~\cite{Okun,Holdom,Foot:1991kb}.
Here, $j^{\mu}$ is an external ordinary electromagnetic current, $A^{\mu}$ is the vector potential and $F^{\mu\nu}$ is the electromagnetic field strength tensor. For later use, we also explicitly include the photon self-energy $\Pi^{\mu\nu}$ induced by a potential medium, with a transversal component $\pi_T$ and a longitudinal component $\pi_L$.\footnote{In principle, the medium contribution, of course, arises from the interaction $j_{\mu}A^{\mu}$ with the SM particles. However, it is often convenient to simply include the effective term arising from the medium.} For the dark fields we have $X^{\mu}$, the DP vector potential, and $X^{\mu\nu}$, the DP field strength tensor. We do not consider any dark currents. \\
Note that we also do not consider any mass mixings. In the absence of dark currents, those can be turned into kinetic mixings by a field redefinition.

In the following, we want to generalize the basic equations for observables that have been used to constrain the single DP case. Specifically, we examine the electrostatic potential constrained in Cavendish experiments~\cite{cavendish} (for a review see \cite{Loegl}, current experimental constraints can be found in~\cite{boundary1, boundary2, boundary3}), the photon detection rate in light-shining-through-walls (LSW) experiments~\cite{Okun,Popov,Anselm:1985obz,VanBibber:1987rq,Anselm:1987vj,ALPSII:2025eri} (cf. the reviews ~\cite{Jaeckel:2010ni,LSWRedondo,LSW,Graham:2015ouw,Irastorza:2018dyq}) and the DP contribution to the solar energy loss \cite{Popov,Okun,redondo08,redondo13,Pradler}. 

\subsection{Preliminary steps}\label{sec::2.1}
In this work, we will restrict ourselves to tree-level computations, thus all relevant expressions can be derived from the corresponding equations of motion. Choosing Lorenz gauge for the photon field, the equations of motion for the different values of the spacetime index $\mu$ decouple and we suppress it in the following. The equations of motion read,  
\begin{gather}\label{eq::2}                
    \left[
    \Box
    \begin{pmatrix}
        1 & \epsilon_1 & \epsilon_2 & ... \\
        \epsilon_1 & 1 & \chi_{12} &\\
        \epsilon_2 & \chi_{12} & 1 \\
        ...\\
    \end{pmatrix}
    + 
    \begin{pmatrix}
    \pi_a & 0 & 0 & ...\\
    0 & m_1^2 & 0 \\
    0 & 0& m_2^2 \\
    ...\\
    \end{pmatrix}
    \right]
    \begin{pmatrix}
        A \\ X_1 \\ X_2\\...
    \end{pmatrix}
    =
    \begin{pmatrix}
        j \\ 0 \\ 0\\...
    \end{pmatrix}\,.
\end{gather}
Here, $\pi_a$ denotes either the transverse or longitudinal photon self-energy contribution in a medium, depending on the considered components. 

To solve these equations, it helps to remove the photon-DP mixing.  
To this end, we carry out a basis transformation that separates the DP and photon contributions, analogously to the single DP calculation~\cite{Holdom}. The transformation matrix is given by the identity, with the first row vector replaced by 
$\begin{pmatrix}
    1 & -\epsilon_1 & -\epsilon_2 ...
\end{pmatrix}$ 
$\equiv $
$\begin{pmatrix}
    1 & -\Vec{\epsilon}~^T
\end{pmatrix}$
, which is multiplied from the right side. Multiplying from the left side with its transposed then gives (up to first order in the $\epsilon$),
\begin{gather}\label{eq::3}                
    \left[
    \Box
    \begin{pmatrix}
        1 & 0 & 0 & ... \\
        0 & 1 & \chi_{12} &\\
        0 & \chi_{12} & 1 \\
        ...\\
    \end{pmatrix}
    + 
    \pi_a
    \begin{pmatrix}
    1 & -\Vec{\epsilon}~^T\\
    -\Vec{\epsilon} & \Vec{\epsilon}\Vec{\epsilon}~^T
    \end{pmatrix}
    +
    \begin{pmatrix}
    0 & 0 & 0 & ...\\
    0 & m_1^2 & 0 \\
    0 & 0& m_2^2 \\
    ...\\
    \end{pmatrix}
    \right]
    \begin{pmatrix}
        A' \\ X_1' \\ X_2'\\...
    \end{pmatrix}
    =
    j
    \begin{pmatrix}
        1 \\ -\vec{\epsilon}
    \end{pmatrix}\,.
\end{gather}

The kinetic matrix is symmetric, hence it can be diagonalized by an orthogonal redefinition of the DP field basis and subsequently transformed into the identity by field renormalizations. The symmetry of all the other matrices will be conserved and thus the mass matrix can be diagonalized in the same way. The mass matrix and the matrix factorized by $\pi_a$ in \eqref{eq::3} will see their structure preserved, though the exact coefficients differ. We call those the diagonalized or effective DP parameters $\epsilon^*$ and $m^*$. If $\pi_a=0$, this already fully decouples our N+1 equations,
\begin{gather}\label{eq::decoupled}                
    \left[
    \Box
    \begin{pmatrix}
        1 & 0 & 0 & ... \\
        0 & 1 & 0 &\\
        0 & 0 & 1 \\
        ...\\
    \end{pmatrix}
    +
    \begin{pmatrix}
    0 & 0 & 0 & ...\\
    0 & {m_1^2}^* & 0 \\
    0 & 0& {m_2^2}^* \\
    ...\\
    \end{pmatrix}
    \right]
    \begin{pmatrix}
        A'' \\ X_1'' \\ X_2''\\...
    \end{pmatrix}
    =
    j
    \begin{pmatrix}
        1 \\ -\vec{\epsilon*}
    \end{pmatrix}\,.
\end{gather}
The procedure sketched here is carried out more explicitly in Appendix~\ref{AppendixA}. There, also the full expression with non-vanishing $\pi_a$ can be found.

For some experimental situations, notably LSW, it is convenient to use a basis where the current is simple. To this end, we can partially reverse the first basis redefinition by exchanging the $\epsilon_i$ in the transformation matrix with the effective couplings $\epsilon^*_i$ and we finally arrive at, 
\begin{gather}\label{eq::4}                
    \left[
    \Box
    \begin{pmatrix}
        1 & \epsilon^*_1 & \epsilon^*_2 & ... \\
        \epsilon^*_1 & 1 & 0 &\\
        \epsilon^*_2 & 0 & 1 \\
        ...\\
    \end{pmatrix}
    +
    \begin{pmatrix}
    \pi_a & 0 & 0 & ...\\
    0 & {m^*_1}^2 & 0 \\
    0 & 0& {m^*_2}^2 \\
    ...\\
    \end{pmatrix}
    \right]
    \begin{pmatrix}
        A''' \\ X'''_1 \\ X'''_2\\...
    \end{pmatrix}
    =
    j
    \begin{pmatrix}
        1 \\ 0 \\ ..
    \end{pmatrix}.
\end{gather}

Depending on the physical situation, the parameters can be plugged into  Eq.~\eqref{eq::decoupled} or Eq.~\eqref{eq::4} and subsequently the solutions to the leading order in $\epsilon$ can be obtained straightforwardly, or can even be guessed from the single DP setting. The derivation of the experimental observables then proceeds analogously to the single DP setting.

\subsection{Cavendish experiments}
Cavendish experiments~\cite{cavendish,Loegl, 
boundary1, boundary2, boundary3}  are a powerful tool to search for deviations from the $1/r$ behavior expected in ordinary electrodynamics with a massless photon. 
They exploit that it is a unique feature of a $1/r$ potential that the interior of a conducting surface is field-free, i.e., the potential is constant and there is no electric field or non-vanishing voltage. They are sensitive to deviations from the $1/r$ behavior as provided by a single or multiple (massive) DPs~\cite{Popov}.

For a Cavendish experiment, we can solve Eq.~\eqref{eq::4} (with $\pi=0$). The matrix equation can then be fully diagonalized by applying analogous transformations to those leading to \eqref{eq::3} with $\epsilon_i$ replaced by $\epsilon_i^*$. 

The electromagnetic potential is then found by evaluating the decoupled equations for the current density of a point charge and integrating the resulting potentials over the spherical shells in the Cavendish setup. We find the expected Yukawa terms for the DP fields in the presence of a point charge,
\begin{eqnarray}\label{extra::1}
    A(r)&=&\frac{q}{4\pi r} \\
    X_i(r)&=&\epsilon_i^*\frac{q}{4\pi r}\exp{(-m_i^* r)}.
\end{eqnarray}

Ultimately, we are interested in the potential energy for a charged SM particle. This is given by the scalar product of the first unit vector multiplied by q and the field vector in our initial field basis. Applying basic linear algebra, we can quickly check that this is equal to,
\begin{align}\label{extra::2}
   q
   \begin{pmatrix}
        1 & \epsilon_1^* ..
    \end{pmatrix}
    \begin{pmatrix}
        A'' \\ X''_i \\ ..
    \end{pmatrix}\,.
\end{align}
Overall, this corresponds to a sum of the different dark photon contributions to the potential energy between two charged particles. 

Cavendish experiments usually feature charged shells between which the potential is measured. We therefore have to integrate over the spherical shells. Doing this yields the voltage ratio between the inner and the outer shells as described in~\cite{Loegl},
\\
\begin{equation}\label{eq::5}
     \frac{U_{ab}}{U_{cd}}\approx\sum^n_{i=1}{\epsilon_i^*}^2\frac{1}{2m_i^*}\frac{(e^{-m_i^*d}/d+e^{-m_i^*c}/c)\times(\sinh{(m_i^*a)}/a-\sinh{(m_i^*b)}/b)}{1/c-1/d}.
 \end{equation}
\\
 Where a, b, c and d are the radial distances of the charged shells. The equation makes it evident that the single DP contributions simply add up constructively in the overall voltage ratio. \footnote{As the sum only contains squares of the mixing, this holds independent of the sign of the $\epsilon_i^*$. One can understand this by noting that this sign can simply be removed by a field redefinition of the DP field with a minus sign.} 
 
 The experimental quantity constrained by Cavendish experiments $\gamma_\text{exp}^\text{Cav}$ is the voltage ratio. For this analysis, we use the results from~\cite{boundary3} and impose \begin{equation}\gamma_\text{exp}^\text{Cav}=\frac{U_{ab}}{U_{cd}} \leq 8.0\times 10^{-15}.
 \end{equation}

 \subsection{LSW experiments}\label{sec:2.3}
 Dark photons can also be searched for with light-shining-through-walls experiments~\cite{Okun,Anselm:1985obz,VanBibber:1987rq, Anselm:1987vj,Popov,ALPSII:2025eri} (for some reviews see, e.g.~\cite{Jaeckel:2010ni,LSWRedondo,LSW,Graham:2015ouw,Irastorza:2018dyq}). (Explicitly, we will use the most recent result by the ALPS II collaboration~\cite{ALPSII:2025eri}.) They exploit that kinetic mixing of the photon with massive dark photons leads to a misalignment of the interaction and propagation eigenstates, resulting in the photon sector analogue of neutrino oscillations. As the “wall" is constructed to filter out any ordinary photon component, it allows only ``dark'' states to pass through. These can then oscillate back into the ``visible'' state and may consequently be detected. This measurement is essentially free of any Standard Model background (see~\cite{Raffelt:1987im,Gies:2009wx} for some small exceptions).

Turning to the calculation in the case of multiple dark photons. In the propagation volume of LSW experiments, both the external current and the photon self-energy vanish. 
However, as production, projection by the wall and detection at the end are determined by the interaction with ordinary matter, it is convenient to use the interaction basis, i.e. Eq.~\eqref{eq::4}.

 After diagonalizing \eqref{eq::4} it is clear that the solutions 
 are plane waves with a dispersion relation $\omega^2-k_i^2={m^*_i}^2$.
 We start with an initially purely photon-like interaction state\footnote{For a detailed reasoning why this is expected, see Appendix B of~\cite{LSW}.} that spatially evolves up to the point of the opaque wall, where it is projected onto its non-interacting parts. Inside the wall, we assume the electromagnetic interaction eigenstate to receive a large, effective and potentially complex mass. From~\eqref{eq:app.EOMlargeAmass} it is then readily understood that the interaction and mass eigenstates agree, such that the DP fields propagating up to the end of the wall fulfill the dispersion relation $\omega^2-k^2={m_i^*}^2$. Effectively they do not interact with the wall. But for different masses, propagation inside the walls changes the relative phases between the DPs. Afterwards, the system is once again being spatially evolved up to the position of the detector. At this point, we take the absolute value squared of the scalar product between the interacting state and our spatially evolved state to obtain the detection probability and thus the rate of detection.

 In a tedious but straightforward calculation, we find the following result for the LSW probability,
\begin{eqnarray}\label{eq::6}
P_{\gamma\rightarrow\gamma'\rightarrow\gamma}
\!\!&=&\!\!16\sum_{i,j=1}^n{\epsilon_i^*}^2{\epsilon_j^*}^2 \sin\left(\frac{{m_i^*}^2L}{4\omega}\right)\sin\left(\frac{{m_i^*}^2L'}{4\omega}\right)
\\\nonumber
&&\qquad\qquad\qquad\times
\sin\left(\frac{{m_j^*}^2L}{4\omega}\right)
  \sin\left(\frac{{m_j^*}^2L'}{4\omega}\right)\cos\left(\frac{({m_i^*}^2-{m_j^*}^2)(L+L'+2L_w)}{4\omega}\right)\,,
\end{eqnarray}
where $\omega$ is the frequency of the laser employed in the experiment, $L$ and $L'$ are the propagation lengths in front and behind the wall, respectively, and $L_w$ is the wall length. Note that we also used $m_i^* \ll \omega$ to simplify the expression. 

In contrast to the Cavendish result, the conversion probability contributions of the individual DPs do not simply add up, but can interfere destructively\footnote{Cf. also~\cite{deGiorgi:2025ldc} for the case of ALPs.}. Thus, an additional DP may decrease the particle rate we measure for a given length and frequency setup. In fact, it is possible to calculate the effect of the $n$-th DP using standard trigonometric identities
\begin{align}\label{extr::4} 
\Delta P_{\gamma\rightarrow\gamma'\rightarrow\gamma}=&\sum_{i,j=1}^{n}p_{ij}-\sum_{i,j=1}^{n-1}p_{ij}
\\=&16\Lambda{\epsilon_n^*}^2\sin\left(\frac{{m_n^*}^2L}{4\omega}\right)\sin\left(\frac{{m_n^*}^2L'}{4\omega}\right)\sin\left(\frac{{m_n^*}^2(L+L'+2L_w)}{4\omega}+\phi\right)\nonumber\\&+{\epsilon_n^*}^4\sin\left(\frac{{m_n^*}^2L'}{4\omega}\right)^2 \sin\left(\frac{{m_n^*}^2L}{4\omega}\right)^2\,,
\end{align}
where $p_{ij}$ are the summands in \eqref{eq::6} and $\Lambda$ and $\phi$ are calculable constants (based on the other $n-1$ $m_i^*$ and $\epsilon_i^*$). Note that the last term plays an increasingly subleading role for large $n$, because $\Lambda$ contains a sum over the other DP species. As we deal with three sinusoidal functions, the zeros of the individual sinusoidals can be extracted to determine the respective mass ranges where the $n$-th contribution is negative or positive.
Whether these contributions are macroscopically positive or negative will be explored in the statistical part of this work. 

The oscillation probability in Eq.~\eqref{eq::6} is the quantity constrained by LSW experiments, the limit based on the results of~\cite{ALPSII:2025eri} is, 
\begin{equation}  \gamma_\text{exp}^\text{LSW}=P_{\gamma\rightarrow\gamma'\rightarrow\gamma}\leq 2.6 \times 10^{-26}.  
\end{equation}

\subsection{Solar energy loss rate}
In the presence of light, $\lesssim 10\,{\rm keV}$, feebly interacting particles, stars can suffer a significant extra energy loss, see~\cite{Raffelt:1996wa} for a textbook. 
Observations of stars, most prominently the sun, therefore lead to powerful constraints on the existence of such particles, notably dark photons~\cite{Popov,Okun,redondo08,redondo13,Pradler}. 

Here, we focus on the sun.
To quantify the solar energy loss rate caused by multiple dark photons, we follow the approach of \cite{redondo08}, i.e., we approximate the “exit probability" for a DP to leave the sun by projecting the photon interaction-state, which we assume to be created in the sun, on the mostly DP-like propagation eigenstates. We distinguish between the DP energy loss originating from the T- and L-channel, respectively. There are no external currents. However, we have a non-zero photon self-energy since we consider the sun to be a dense medium.\footnote{As already mentioned, on a fundamental level, the photon-self energy is created by the interactions of photons with the ordinary electromagnetic current.} Explicitly we have (on-shell)~\cite{redondo08,redondo13,Pradler}:
\begin{eqnarray}\label{eq::7}
    \pi_T&=&\omega_P^2-i\omega \Gamma \qquad\qquad\qquad\, \text{Transversal photons} \\
    \pi_L&=&\omega_P^2-k^2-i\omega \Gamma  \qquad\qquad\text{Longitudinal photons}\,.
\end{eqnarray}
Here, $\omega_P$ is the plasma frequency, $\omega$ and $k$ are the particle's energy and the absolute value of its momentum. $\Gamma$ is the rate at which thermal equilibrium is reached, which is connected to the imaginary part of the photon self-energy by $\text{Im}({\pi})=-\omega \Gamma$. For a review of these quantities see, again, \cite{Raffelt:1996wa}. Relevant derivations of the quantities involved are carried out in \cite{weldon,braaten}. It can be expressed by the production rate of photons in the sun $\Gamma=\Gamma_\text{prod}(e^{\omega/T}-1)$ (whose dominant contributions are inverse Bremsstrahlung and Compton-scattering. We consider the same expressions and thus work at the same level of precision as \cite{redondo13}). 

A crucial aspect of this is that the dispersion relation for longitudinal photons in \eqref{eq::7} is only valid on-shell, as it is, in general, energy-dependent. As has been discussed in~\cite{redondo13}, this can be fixed by renormalizing the couplings $\epsilon^*_i \rightarrow \sqrt{Z_L}\epsilon^*_i$ with $Z_L=\omega^2/K^2$, where K is the 4-momentum. Neglecting this will lead to an incorrect proportionality between the energy loss rate and the DP mass in the small mass regime \cite{Pradler}. Following the same approach, we obtain the differential energy loss rate per unit volume by multiplying the aforementioned exit probability with $\omega$ and the photon production rate,
\begin{eqnarray}
\label{eq::8}
\frac{dW_T}{d\omega dV}&=&\sum_{i=1}^n \frac{1}{\pi^2}\frac{\omega^2\sqrt{\omega^2-m_X^2}}{e^{\omega/T}-1}\frac{{\epsilon_i^*}^2{m_i^*}^4}{(\omega_P^2-m_X^2)^2+(\omega \Gamma)^2}\Gamma_\text{prod}(\omega,\vec{r}) ,
\\ 
\label{eq::8b}
\frac{dW_L}{d\omega dV}&=&\sum_{i=1}^n \frac{1}{2\pi^2}\frac{\omega^4\sqrt{\omega^2-m_X^2}}{e^{\omega/T}-1}\frac{{\epsilon_i^*}^2 {m_i^*}^2}{(\omega_P^2-\omega^2)^2+(\omega \Gamma)^2}\Gamma_\text{prod}(\omega,\Vec{r}).    
\end{eqnarray}

We conclude that the energy loss rates in the L- and T-channels are additive with respect to multiple different DPs. Naturally, this also applies to the several approximative expressions valid in specific mass ranges derived in \cite{redondo08,redondo13}. The overall luminosity can then be obtained by integrating \eqref{eq::8} over a solar model. Explicitly, we use the one from~\cite{Serenelli}. 

In order to deduce limits on the kinetic mixing, one customarily imposes that the solar energy loss, 
\begin{equation}
W=\int^{\infty}_{0} d\omega \int_{\rm{V_{\rm sun}}} dV \frac{dW}{d\omega dV},
\end{equation}
whose observational limit we define as $\gamma_\text{exp}^\text{Solar}$, via the DP channel does not exceed 10\% of its actual luminosity~\cite{redondo08}. Using the solar model~\cite{Serenelli}, this corresponds to 
\begin{equation}\gamma_\text{exp}^\text{Solar}=W \leq 1.6\times10^{29} \text{ eV}^2.
\end{equation}

\section{Numerical/statistical analysis}\label{sec::3}
With the expressions derived in the previous section, it is possible to restrict the parameters of arbitrarily many DPs by comparing their physical signatures to the limits obtained from experiments and observations. However, the bounds in the high-dimensional parameter space are fairly difficult to visualize and interpret.
Therefore, we perform a statistical analysis, where we quantify the overall size of masses and mixings with a small number of scale parameters and take the detailed values to be statistically distributed. 

Explicitly, we define a single mass scale ($M$) and coupling ($A$) parameter, respectively, and describe the individual dark photon parameters as random variables, whose probability distribution functions (PDFs) depend on $A$ and $M$. This then renders the effect in each of the physical settings a random variable with a corresponding distribution. We constrain the pairs ($A$,$M$) by checking whether the median of these distributions exceeds any of the experimental thresholds. Thus, a combination of $A$ and $M$ is taken to be “ruled out" if less than 50\% of their associated parameter sets or, equivalently, 50\% of the models
agree with the experimental results.

Naturally, the choice of the percentage of models we require to be in agreement with experimental evidence is somewhat arbitrary. One could equally well be more rigid and, e.g., require 95\% compatibility to keep a pair of ($A$,$M$). To illustrate the dependence on this, in the later Figs.~\ref{extrafig::1} and \ref{extrafig::2} we exemplarily examine this question by showing the overlay of the limits of 1000 random models. 

To obtain concrete results, we need to specify a distribution for the random parameters. While there are in principle many possibilities, for simplicity, we limit ourselves to the following equiprobabilistic distributions for the specified powers of the parameters.
\begin{align}\label{eq::9}
\mu'(m_1^2, ...,m_n^2)=& \frac{1}{\mathcal{N}}\mathbf{1}_{(\sum_i^n m_i^2/n=M^2)}\\
\mu'(\epsilon_1^p, ...,\epsilon_n^p)=& \frac{1}{\mathcal{N}}\mathbf{1}_{(\sum_i^n \epsilon_i^p=A^p)}. \nonumber
\end{align}
Here, $p$ is a natural number, $\mathbf{1}$ represents the indicator function and $\mathcal{N}$ takes care of the normalization. We thus assign each set of parameters, which fulfill the conditions in the subscript of the indicator function, equal probability. All combinations of (squared) masses which have mean $M^2$ and all combinations of $\epsilon^p$ that add to $A^p$ are weighted evenly\footnote{One could, of course, also choose $M^2$ to be the sum of the individual parameters. Effectively, this corresponds to a rescaling with the DP number $N$.}.

Mathematically, Eq.~\eqref{eq::9} describes a random variable with a Dirichlet distribution with order parameter $n$, and concentration parameters uniformly set to 1, c.f.~\cite{zacks2013examples}. 
In the following, we interpret the equations describing the physical observables as functions of random variables and use the PDF in \eqref{eq::9} to calculate their expectation values and variances. This gives us a macroscopic idea of the magnitude of the physical effect induced by multiple DPs, whose couplings are around the order $A$ and whose masses accumulate to $M$. 

\bigskip

Before going into a full numerical analysis, it is nice to get some analytical insight. To do so, we will consider in Sec.~\ref{sec::3.1} a simplified situation where we assume the mixing parameters $\chi_{ij}$ to be 0. I.e., we directly attach the probability distributions to the diagonalized parameters in Eq.~\eqref{eq::4}. 
As discussed above, we aim to place our limits based on the percentage of models excluded and hence consider the median. However, to derive analytical expressions, it is much easier to use the average. That said, we find that, for the most part, the average limits do not deviate substantially from those obtained via the median. We comment where deviations occur.

Then, in subsection~\ref{sec::3.2}, we consider the full picture.
To do so, we also need to assign a probability distribution to the random variables $\chi_{ij}$. Similarly to the other parameters, we use the following simple choice,
\begin{equation}\label{eq2::1}
    \mu'(\chi_{12},...)=\frac{1}{\mathcal{N}}\mathbf{1}_{[-\chi_{\rm{max}}/2,\chi_{\rm{max}}/2]},
\end{equation}
where the same notation as in~\eqref{eq::9} was used and $\chi_\text{max}$ is an arbitrary (but in practice $<\mathcal{O}(1)$) number. 

\subsection{Simple special case: diagonal mixing matrix}\label{sec::3.1}

As a warm-up exercise, and to get some analytical insights, we look at the setting described in the previous section, but with all $\chi_{ij}$ set to 0. Then the effective couplings and masses in Eqs.~\eqref{eq::5}, \eqref{eq::6}, \eqref{eq::8} and \eqref{eq::8b} are directly given by the random variables $\epsilon_i$ and $m_i$. 

This case is particularly simple, as, by assumption, the $\epsilon_i$ and $m_i$ are independent of each other. Statistical moments such as the mean and the variance of the various observables can be computed as the integral of powers of the relevant observables multiplied by the PDFs ~\eqref{eq::9}. These integrals, which are essentially sums of products between polynomials of $\epsilon^p_i$ and functions of $m^2_i$, can be split into two parts. Since the individual summands in Eqs.~\eqref{eq::5}, \eqref{eq::6}, \eqref{eq::8} and \eqref{eq::8b} do at most depend on two different mass and coupling parameters at a time, the integral over all the other parameters can be carried out straightforwardly, the PDF of the remaining ones is then described by,
\begin{align}\label{eq::9+1}
\mu'(m_i^2)=& \frac{1}{M^2}\left(1-\frac{m_i^2}{M^2n}\right)^{n-1}\mathbf{1}_{(m_i^2<n\times M^2)} \\
\mu'(m_i^2, m_j^2)=&\frac{(n-1)}{M^4n} \left(1-\frac{m_i^2+m_j^2}{M^2n}\right)^{n-2}\mathbf{1}_{(m_i^2+m_j^2<n\times M^2)} \nonumber\\ 
\mu'(\epsilon_i^p)=& \frac{n}{A^p}\left(1-\frac{\epsilon_i^p}{A^p}\right)^{n-1}\mathbf{1}_{(\epsilon_i^p< A^p)} \nonumber\\
\mu'(\epsilon_i^p, \epsilon_j^p)=& \frac{n(n-1)}{A^{2p}}\left(1-\frac{\epsilon_i^p+\epsilon_j^p}{A^p}\right)^{n-2}\mathbf{1}_{(\epsilon_i^p+\epsilon_j^p< A^p)} \,.\nonumber
\end{align}

With this setup in place, it is possible to draw limit plots similar to the traditional ones in the single-DP setting. To achieve this, we generate effective couplings and masses for different $M^2$, whereas $A$ is set to 1 in the PDF \eqref{eq::9} (alternatively, the generated $\epsilon_i$ can be divided by $A$, which corresponds to a new random variable x). Performing this many times (N=1000) and taking the expectation value E[x] yields 
the average observable for a mixing of size 1. In the present context, 
we then consider a value of $A$  
to be permissible if the mean
physical effect associated with it remains below the experimental sensitivity, 
\begin{equation}\label{eq::boundscaling}
    A\leq \left(\frac{{\rm exp.\,\,limit}}{{\rm E[theoretical \,\,result \,\,for\,\,} A=1] }\right)^{k}=\left(\frac{\gamma_\text{exp}}{\gamma_\text{th}}\right)^k.
\end{equation}
Here, the power $k$ is given by the power of the $\epsilon_{i}$ that appears in the calculation of the observable.
The experimental and observational limits are taken from~\cite{boundary3,ALPSII:2025eri,redondo08,Serenelli} and for the solar luminosity, we have employed an interpolation with respect to the mass of the integrals of Eqs.~\eqref{eq::8} and \eqref{eq::8b}. This speeds up the calculation.

\begin{figure}[!t]
    \centering
    \includegraphics[width=0.48\linewidth]{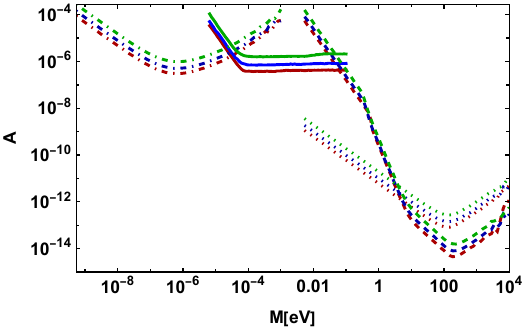}
    \includegraphics[width=0.48\linewidth]{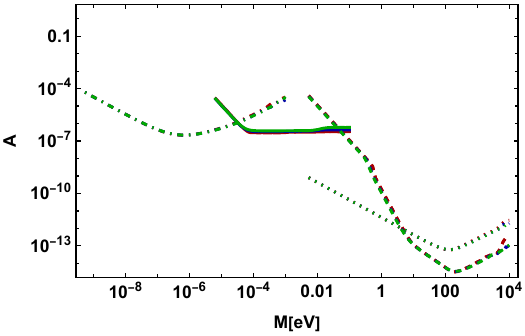}
    \caption{Average bounds depending on the scaling parameter $A$ from Cavendish (leftmost curve) and LSW (centre curve) experiments, and DP contribution to the solar luminosity (rightmost curves). We assume no DP mixing. Green: $n=40$, blue: $n=10$, red: $n=3$. Dot-dashed: Cavendish, continuous: LSW, dashed: Solar lifetime (transverse), dotted: Solar lifetime (longitudinal). Left panel: $p=1$, right panel: $p=2$. Sample size N=1000.}
    \label{fig::1}
\end{figure}

The results are shown in Fig.~\ref{fig::1}. 
An important qualitative takeaway is that the $M$-dependence of the average limits
in the low mass range is identical to the $m$-dependence of the single DP limits in that range. In contrast,  
in the high mass regions, the bounds typically weaken with a power of $M$ but not exponentially, as would be the case in many situations in the single dark photon case. 
The reason is essentially that we average the signal. As follows from the used distributions,  
there is a power-like probability for at least one of the dark photons to be below the threshold for the onset of exponential behavior, therefore giving a contribution to the (average) experimental signature that is only suppressed by a power law itself. We will see in the following that this behavior also differs from the one we will observe when considering the median instead of the average. With the median, the limits will again fall faster for high masses.
Due to the averaging, the LSW bounds do not display any oscillatory behavior. This, too, is in contrast to their single DP non-averaged counterparts. 

Moreover, the LSW bounds in Fig.~\ref{fig::1} also slightly change around $0.01~\text{eV}$. At this point the wall thickness, which we take to be $1~{\rm cm}$\footnote{We would like to thank A.~Lindner for the relevant information on the ALPS II setup.}, becomes relevant. In this situation the relative phases between the different DP species acquired inside the wall become relevant and destroy the remaining coherence. \footnote{This is not entirely obvious from  Eq.~\eqref{eq::6}. However, on closer inspection the sums in Eq.~\eqref{eq::6} can be written such that there are terms which are $\sim \cos\left(\frac{L_{w}}{2\omega}({m_{i}^{*}}^2-{m_{j}^{*}}^2)\right)$ which average out for large mass differences. Note that a trigonometric decomposition of the non-coherent contributions shows that they do not average out.} 
 
From Fig.~\ref{fig::1} we can also see that the strength of the limits depends on the number of DPs, but the direction of this may be affected by the choice of distribution.
To get a clearer picture, let us get some analytical insight into the dependence of the limits as well as the width of the distribution on the number of dark photons $n$. 

At $n\to \infty$ analytical results can be obtained for the statistical moments of \eqref{eq::5}, \eqref{eq::6}, \eqref{eq::8} and \eqref{eq::8b}. This becomes apparent when one realizes that the PDF describing the masses \eqref{eq::9+1} weakly converges to an exponential distribution in the $n \rightarrow \infty$ limit. In particular, it is independent of the dark photon number $n$~\footnote{This is, of course, a direct consequence of our normalization in \eqref{eq::9}.}. Explicitly taking the limit in (\ref{eq::9+1}) (or rather the CDF to prove weak convergence), we find,
\begin{align}\label{extra::5}
\mu_{n\rightarrow \infty}(m_i^2)=& \frac{1}{M^2}\exp\left(-\frac{m_i^2}{M_i^2}\right) \\
\mu_{n\rightarrow \infty}(m_i^2, m_j^2)=&\frac{1}{M^4}\exp\left(-\frac{(m_i+m_j^2)}{M_i^2}\right). \nonumber    
\end{align}

Consequently, the leading order $n$-behavior of any statistical moment is solely determined by the integral of the couplings.\footnote{The mass-dependent integral factorizes as an asymptotically $n$-independent coefficient.} The expression for the observables' means and variances can be interpreted as weighted sums of the squared couplings (or the square of this sum in the coherent regime of the LSW detection rate). In terms of the $\epsilon_i$, the expressions for the physical effects in all our constraining settings are merely polynomials, hence the integral by the measure \eqref{eq::9+1} can be computed straightforwardly to any order in $n$. In the high- and low-mass range, it is even possible to give explicit expressions for the average and the variance of the different observables as a function of $M$, $A$ and $n$. To this end, the exponent in \eqref{extra::5} is Taylor expanded in the $m_{i}$ for the small $M$ limit and the integral is cut off by some value $M$ that lies past the bulk of the integrand in the high $M$ limit. Hence, for large $M$, the only signature of $M^2$ is in the prefactor, as the $m_i$ integrals converge to a constant. This explains the polynomial weakening of the bounds alluded to earlier.
The resulting expressions to leading order in $n$ for the maximum permissible value of $A$ given a certain $M$ are thus explicitly calculable using~\eqref{eq::boundscaling} and are listed in Tab.~\ref{table::1} and Tab.~\ref{table::2}.

\begingroup
\setlength{\tabcolsep}{10pt} 
\renewcommand{\arraystretch}{1.5} 
\begin{table}[t] 
\centering
  \begin{tabular}{c|ccc}    
    &$p$=1 & $p$=2 & $p$=4 \\
    \hline
    
    Cavendish  &$n^{1/2}$ 2.3\si{\times 10^{-14} \eV\per}$M$ & 3.2\si{\times 10^{-14} \eV\per} $M$ & $n^{-1/4}$ 3.4\si{\times 10^{-14} \eV\per} $M$ \\
    LSW &$n^{1/2}$ 3.3\si{\times 10^{-14}  \eV^2\per} $M^2$ & 4.6\si{\times 10^{-14} \eV^2\per} $M^2$ & $n^{-1/4}$ 4.8\si{\times 10^{-14} \eV^2\per}$M^2$ \\
    
    Sun (T)  & $n^{1/2}$ 6.9\si{\times 10^{-10} \eV^2\per} $M^2$ & 1.0\si{\times 10^{-9} \eV^2\per} $M^2$ & $n^{-1/4}$ 1.0\si{\times 10^{-9} \eV^2\per} $M^2$ \\ 
    Sun (L) &$n^{1/2}$ 9.4\si{\times 10^{-13}\eV\per} $M$& 1.3\si{\times 10^{-12} \eV\per} $M$&$n^{-1/4}$ 1.4\si{\times 10^{-12} \eV\per} $M$\\ 
  \end{tabular}
  \caption{Relation between DP number $n$ and maximum permissible $A$ in the low-mass limit. The maximal $A$ is chosen such that the average of the observables' distribution does not exceed different phenomenological bounds. The expressions were obtained to leading order in $1/n$.}
  \label{table::1}
\end{table}
\endgroup
\begingroup
\setlength{\tabcolsep}{10pt} 
\renewcommand{\arraystretch}{1.5} 
\begin{table}[t] 
\centering
  \begin{tabular}{c|ccc}    
    &$p$=1 & $p$=2 & $p$=4 \\
    \hline
    
    Cavendish &$n^{1/2}$ 1.8\si{\times 10^{-2}} $M/\,\si{\eV}$ & 2.6\si{\times 10^{-2}}$M/\,\si{\eV}$ & $n^{-1/4}$ 2.8 \si{\times 10^{-2}}$M/\,\si{\eV}$ \\
    LSW (coh.) & $n^{1/2}$ 4.1\si{\times 10^{-7}}& 5.9 \si{\times 10^{-7}} &$n^{-1/4}$ 6.1\si{\times 10^{-7}} \\
    LSW (\sout{coh.}) & $n^{3/4}$ 1.7\si{\times 10^{-7}}& $n^{1/4}$ 2.8 \si{\times 10^{-7}} & 3.7\si{\times 10^{-7}} \\
    
    Sun (T)  &$n^{1/2}$ 6.7\si{\times 10^{-18}} $M/\,\si{\eV}$ & 9.5 \si{\times 10^{-18}}$M/\,\si{\eV}$ & $n^{-1/4}$ 1.1 \si{\times 10^{-17}}$M/\,\si{\eV}$ \\ 
    Sun (L)  &$n^{1/2}$ 4.5\si{\times 10^{-17}}$M/\,\si{\eV}$ & 6.4 \si{\times 10^{-17}}$M/\,\si{\eV}$ & $n^{-1/4}$ 6.8 \si{\times 10^{-17}} $M/\,\si{\eV}$ \\   
  \end{tabular}
  \caption{As in Table~\ref{table::1} but in the high-mass limit. The markers ``coh" and ``\sout{coh}" apply to the mass regimes $\Delta m^2 \ll \omega/L_\text{w}$ and $\Delta m^2 \gg \omega/L_\text{w}$ and indicate whether the coherent contributions in Eq.~\eqref{eq::6} play a role and modify the $n$-behavior. }
  \label{table::2}
\end{table}
\endgroup

We remark that the $n$-dependencies in Tabs.~\ref{table::1} and \ref{table::2} are identical in the low- or the high-mass regime (except for the case of LSW experiments which we shall comment on shortly), as predicted in our discussion above. Therefore, the asymptotic behavior for large $n$ in these two respective regimes, is independent of the value for $M$. 

The leading $n$- and $M$-behavior of the variance and the mean of the distributions
are shown in App.~\ref{AppendixB}.  The results in App.~\ref{AppendixB} can be obtained via Eq.~\eqref{eq::boundscaling}.  From there, it can also be understood that the scaling of the relative uncertainty behaves as,
\begin{equation}
\label{eq:variancemean}
    \frac{\textbf{variance}}{\textbf{mean}^2}\sim\frac{1}{n}.
\end{equation}
These results are expected due to the interpretation of the observables as weighted sums of the coupling parameters. If we fix their squared sum for any $n$, it is only natural that their weighted squared sum will also remain constant with respect to $n$. If we work with a constant $p=1$ norm of the coupling vector (which is, of course, bigger than or equal to the 2-norm), we expect the weighted squared sum to decrease with $n$. 
The converse logic applies when fixing orders of $p$ $>$ 2. 

This also answers the question of whether the DP contributions to the LSW detection rate will statistically add up in a different way than for all the other observables we consider, as, in principle, they can be negative for some $m_n^*$ values. However, we find
that, on average, this observable
behaves analogously to the other ones, where no destructive interference is possible. 
It is noteworthy that the precise answer to this question depends on the mass regime we are looking at. In the low mass regime and the moderately high mass regime ($\Delta m^2 \ll \omega/L_\text{w}$) the coherent terms in Eq.~\eqref{eq::6} are not suppressed. Therefore, the detection probability scales as $\sum ({\epsilon_i^*}^2)^2$. Thus, fixing the 2-norm of the couplings assures that the limits remain constant with respect to $n$ on average. In the very high mass limit however, coherence is no longer observed and the scaling becomes $\sum_i {\epsilon_i^*}^4$. LSW experiments are hence less effective at detecting heavy dark photons than naively assumed. Therefore, when fixing the 2-norm, the limits will actually loosen. Instead, a constant 4-norm assures the asymptotic $n$-independence of the constraints in this case.

We have compared the curves resulting from the analytical behavior given in Tab.~\ref{table::1} and Tab.~\ref{table::2} with the plots in Fig.~\ref{fig::1} and found excellent agreement. Indeed, the bounds become less and less restrictive for higher $n$ in the case of the $p=1$ normalized datasets, while they remain approximately constant for the $p=2$ normalized dataset. In addition, the conclusions from our qualitative discussion of the bounds' $M$-dependencies are also reflected in the plots shown above. Notably, we observe a similar bound shape to the single DP bounds for lower masses and the peak region of each respective observable, whereas the higher mass range differs either in the steepness of the bound's decline (which is, as one would expect, just a straight line now) or the intensity/presence of oscillations which are averaged over. 

\begin{figure}[!t]
    \centering
    \includegraphics[width=0.48\linewidth]{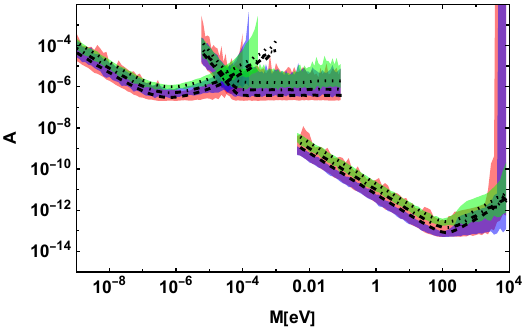}
    \includegraphics[width=0.48\linewidth]{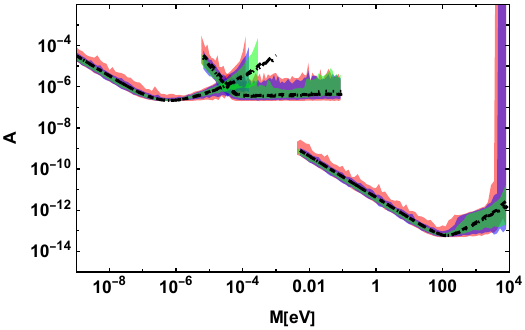}
    \caption{Bounds on mixing parameter $A$ from LSW and Cavendish experiments, and longitudinal photon contribution to the solar luminosity. Colored regions show the whole $A$ range for given $M$, in which at least one is constrained. Green: $n=40$, blue: $n=10$, red: $n=3$. Also shown: maximum $A$ so that the numerical average of observables is reconcilable with observation. We assume no mixing between the DPs. Dotted: $n=40$, dot-dashed: $n=10$, dashed: $n=3$. Left panel: $p=1$, right panel: $p=2$; Sample size N=1000. Note that the transversal luminosity bounds are not shown to simplify the figure. However, they do not bear any surprises.}
    \label{extrafig::1}
\end{figure}

In Fig.~\ref{extrafig::1}, we go beyond just showing the mean and explore the range of limits for random parameter sets. Essentially, the upper line corresponds to the value of $A$ such that none of the generated $1000$ models would be allowed by the experiment/observation, whereas the lower boundary would follow from the requirement that all generated models are experimentally allowed.

\begin{figure}[t!]
    \centering
    \includegraphics[width=0.48\linewidth]{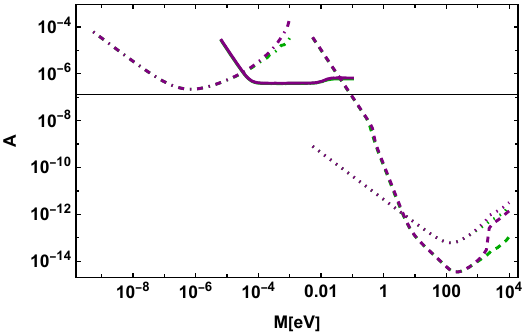}
    \caption{Maximum scale parameter $A$ for given mass $M$ such that median (purple) and mean (green) of the considered observables do not exceed experimental limits. Distribution of couplings and masses governed by \eqref{eq::9}, dark photon mixing set to 0, $p$=2. Sample size N=1000; Dot-dashed: Cavendish, continuum: LSW, dashed: solar lifetime (transverse), dotted: solar lifetime (longitudinal). $n$=40}
    \label{extrafig::4}
\end{figure}

Let us now return to the question of how the mean-based bounds of Fig.~\ref{fig::1} compare to those based on the median. As highlighted for a selection of examples with $p=2$ and $n=40$ in Fig.~\ref{extrafig::4}, they are in good agreement in the lower mass range. For high masses, however, the two plots differ distinctly. The bounds based on the median essentially follow the single dark photon bound shape, while the mean-based bounds decay polynomially in $M$ as pointed out previously. This can, e.g., be seen at the large mass end of the solar T channel curve. The purple line follows the single DP bounds and would thus decay exponentially for even higher masses (which are not included in the plot due to numerical limitations), whereas the green line depletes in a straight manner from the peak on.

As we will see in the following, the (partial) agreement between mean- and median-based constraints is somewhat special to the diagonalized case where we assume no mixing between the DPs (cf. Fig.~\ref{extrafig::4}). Hence, for the more general case, including mixings, we will limit ourselves entirely to a numerical analysis of the median.

\subsection{General mixing}\label{sec::3.2}
Considering the already diagonalized setup is somewhat artificial. It seems more natural to generate all possible mixings and perform the diagonalization explicitly.

Allowing for DP kinetic mixing, the situation becomes more complex. Not only are there more degrees of freedom, notably the mixings $\chi_{ij}$, but in order to arrive at the expressions \eqref{eq::5}, \eqref{eq::6}, \eqref{eq::7}, \eqref{eq::8} and \eqref{eq::8b}, the $n+1$ dimensional equations of motion need to be diagonalized. Thus, obtaining analytical results seems difficult for a larger number of DPs. Hence, we perform a fully numerical analysis.

\begin{figure}[t!]
    \centering
    \includegraphics[scale=1.2]{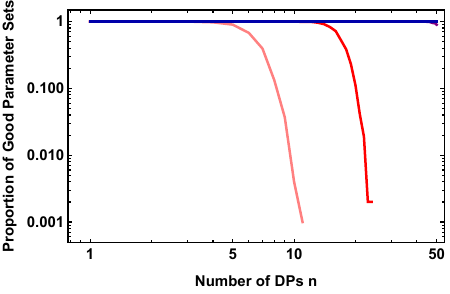}
    \caption{Number of “good” parameter sets as a function of DP number, double logarithmic plot. Colors correspond to different max. $\chi_\text{max}$ values as defined in Eq.~\eqref{eq:chimax}. Blue: 0.05, purple: 0.125, red: 0.25, pink: 0.5. Sample size N=1000.}
    \label{fig::2}
\end{figure}

For the mixing matrix between the different DP photon spaces we again need a distribution. We use Eq.~\eqref{eq2::1},
\begin{equation}
\label{eq:chimax}
    \mu'(\chi_{12},...)=\frac{1}{\mathcal{N}}\mathbf{1}_{[-\chi_{\rm{max}}/2,\chi_{\rm max}/2]}.
\end{equation}

Generating the mixing matrix with such a randomly generated set, it may have negative eigenvalues. Such matrices feature instabilities and are therefore problematic. Hence, we discard those parameter sets. For fixed $\chi_{\rm{max}}$, the fraction of viable parameter sets decreases dramatically with increasing $n$, as shown in Fig.~\ref{fig::2}.

In the end this is not too surprising since multiple relatively small mixings may effectively add to a relatively large effect in the mixing matrix, large enough to change the sign of an eigenvalue.

\begin{figure}[t!]
    \centering
    \includegraphics[scale=1.2]{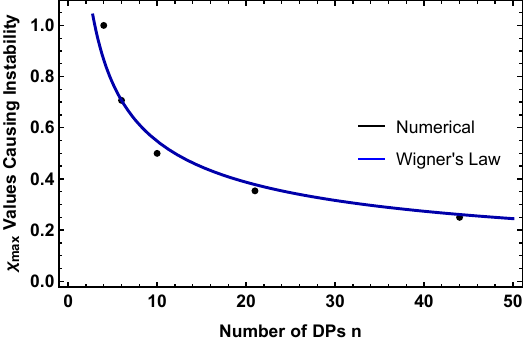}
    \caption{Smallest $\chi_{\rm{max}}$ such that for 1000 samples at least one associated Lagrangian is unstable, i.e., it features at least one negative eigenvalue in the kinetic matrix. We have generated kinetic matrices with  $\chi_{ij}$ distributed according to Eq.~\eqref{eq:chimax}.  Blue curve: Prediction from Wigner's semicircle law (precisely: upper limit for how much eigenvalues of mixing matrix deviate from one).}
    \label{fig::2*}
\end{figure}

Indeed, there is a law behind the number of dark photons at which the Lagrangians start to become unstable. For a given $\chi_\text{max}$, the shift of the lowest eigenvalue of the kinetic matrix from one changes with $n$. This can be seen in two ways. Firstly, it is a direct consequence of Wigner's semicircle law \cite{pastur1972spectrum,wigner1958distribution,tao2010randommatricesuniversalitylocal}, since our kinetic matrix can be described by the sum of an identity matrix and a random matrix that is defined by a random distribution centered around 0 on the diagonals as well as on the off-diagonals. Wigner's semicircle law now states that the largest/smallest eigenvalue scales as $\sqrt{n}$. To be precise, for the $\chi_{ij}$-distribution Eq.~\eqref{eq:chimax}, the largest shift of any eigenvalue from one is given by, 
\begin{equation}
\lambda^\text{EV}_\text{max}=\frac{\chi_{\rm{max}}\sqrt{n}}{\sqrt{3}}.     
\end{equation}
In Fig.~\ref{fig::2*} this is compared to numerical results. 

Alternatively, we can consider a more heuristic argument. In the regime where the smallest eigenvalue is still close to one, it is readily understood that the same is limited from below by the determinant. For sufficiently small off-diagonal contributions, only the quadratic terms in the $\chi_i$ need to be considered, the number of which scale asymptotically as $n^2$. Furthermore, they scale with the order of magnitude of $\chi_\text{max}^2$. Since the contributions to the determinant can be negative or positive, one could describe the evaluation of the determinant as a random walk with respect to the number of contributions. Thus, the “distance" from the initial value grows with $\sqrt{n^2}=n$. Accordingly, in the regime $\chi_\text{max}\ll 1$, we expect that the shift from one scales with $\chi_\text{max}^2\times n $. 

This trend towards instability with increasing number $n$ of the DPs can be avoided by scaling the maximal $\chi_{ij}$-values by $\sqrt{n}$ as we will do in the following.
\begin{figure}[!t]
    \centering
    \includegraphics[width=0.48\linewidth]{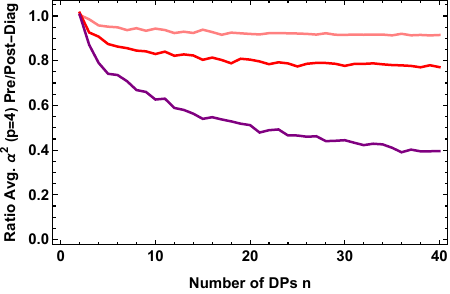}
    \includegraphics[width=0.48\linewidth]{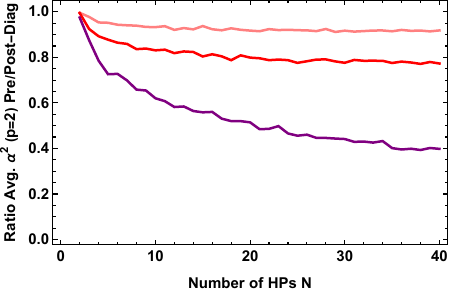}
    \\
    \includegraphics[width=0.48\linewidth]{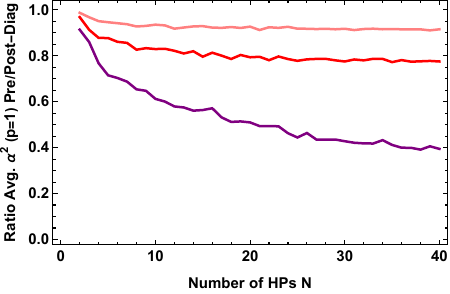}
    \includegraphics[width=0.48\linewidth]{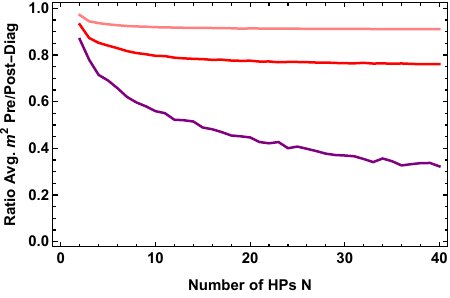}
    \caption{Ratio of the averages ($\frac{\sum_i {\alpha_i^*}^2}{\sum_i {\alpha_i}^2}$/$\frac{\sum_i {m^*}^2}{\sum_i {m}^2}$) varying $\chi_\text{max}\times\sqrt{n}$ (different choices of couplings' powers for normalization $p$=1,2,4). The colors correspond to different $|\chi_\text{max}|\times \sqrt{n}$ values: Pink: 0.5, red: 0.75, purple: 1. Mean calculated with sample size N=1000.}
    \label{fig::4}
\end{figure}
With this choice, we may expect that the effects of non-diagonal kinetic terms will introduce no further strong $n$-dependence. This conjecture is supported by Fig.~\ref{fig::4}, where we compare the quantities $\sum_i{\alpha_i^*}^2$ and $\sum_i{m_i^*}^2$ (hence after diagonalization) with $\sum_i{\alpha_i}^2$ and $\sum_i{m_i}^2$ (the parameters appearing in the undiagonalized Lagrangian), respectively under the assumption that the fourth, second, and first (absolute value) powers of the initial couplings $\alpha$ are fixed. As per our expectation, the ratios do seem to be asymptotically constant with respect to $n$.

\begin{figure}[!t]
    \centering
    \includegraphics[width=0.48\linewidth]{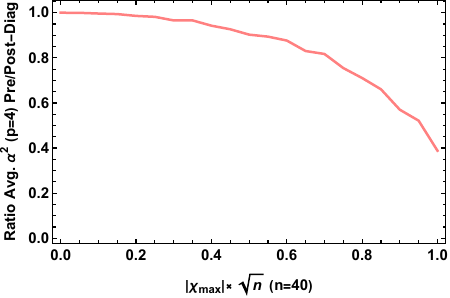}
    \includegraphics[width=0.48\linewidth]{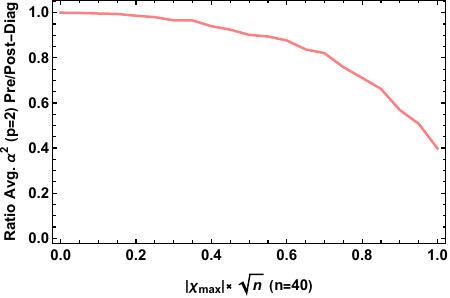}
    \\
    \includegraphics[width=0.48\linewidth]{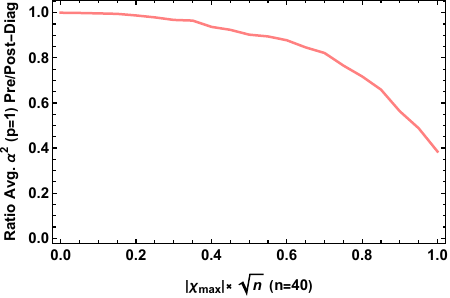}
    \includegraphics[width=0.48\linewidth]{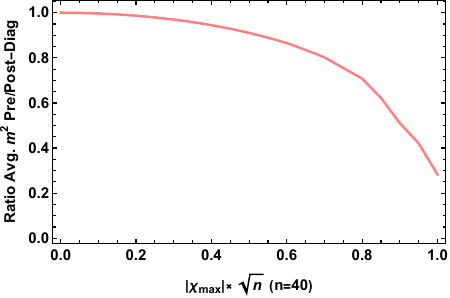}
    \caption{Ratio of the averages ($\frac{\sum_i {\alpha_i^*}^2}{\sum_i {\alpha_i}^2}$/$\frac{\sum_i {m^*}^2}{\sum_i {m}^2}$) varying $\chi_\text{max}\times\sqrt{n}$ (different choices of couplings' powers for normalization $p$=1,2,4). $n=40$ in all plots. The plotted quantity is averaged with the sample size N=1000.}
    \label{fig::4*}
\end{figure}
We are now particularly interested in the cases, where $\chi_\text{max}$ and $n$ are nonetheless large enough to significantly alter the values of the $\epsilon_i$ and $m_i$. Then, the eigenvalues, whose inverses are used for rescaling in the process of diagonalization, can still take values relatively close to 0 and thus distort the scale of individual random variables $m^*_i$ and $\epsilon_i^*$  w.r.t.$~M$ and $A$. In Fig.~\ref{fig::4*}, we show how the ratios of the squared sum of the couplings and masses behave as a function of the parameter $\chi_\text{max}\times\sqrt{n}$, which we shall fix to a particular value for the rest of the analysis. The number $n$ in these figures is fixed at a value, where the curves in Fig.~\ref{fig::4} have already sufficiently converged ($n=40$), whereas only $\chi_\text{max}$ is effectively varied. We find that choosing $\chi_\text{max}\times\sqrt{n}=0.8$ leads to stable but substantial changes to the small mixing case and therefore we choose this value. 

\begin{figure}[t!]
    \centering
    \includegraphics[width=0.48\linewidth]{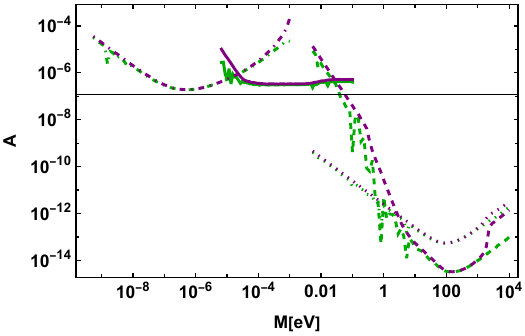}
    \caption{Maximum scale parameter $A$ for given mass $M$ such that numerical median (purple) and mean (green) of the considered observables do not exceed experimental limits.  Case with non-trivial mixings $\chi_{ij}$ between DP species. Distribution of couplings and masses governed by \eqref{eq::9} and \eqref{eq2::1}, $p$=2. Sample size N=1000. Dot-dashed: Cavendish, continuum: LSW, dashed: solar lifetime (transverse), dotted: solar lifetime (longitudinal). $n=40$}
    \label{extrafig::3}
\end{figure}

Let us now turn to the bounds in the ($A$,$M$)-plane. To begin with, let us discuss why we prefer the median compared to the mean. As an example, let us consider the transversal solar photon luminosity in the small mass regime. We find that it exhibits particularly strong variations, since the individual contributions in \eqref{eq::8} scale with the fourth power of $m_i^*$ in this range. Instances with effective masses around \SI{300}{eV} are rare enough to be missed in feasible sample sizes, but are still sufficiently likely to, e.g., affect the true mean value of the distribution due to the mass scaling. This can be seen from the clearly visible fluctuations in the green line of Fig.~\ref{extrafig::3}, where we used a sample of $N=1000$. In some cases, the tails of the distribution render the statistics required to obtain small errors on the mean of the observables prohibitively large. Hence, we prefer the median, which is known to be less prone to outliers.
In principle, the same issue exists even for low values of $M$ in the non-mixed case considered in the previous section, since higher-end tails in the distribution also exist there for low $M$. However, in the absence of mixings between the different dark photons, they are sufficiently suppressed (exponentially for large $n$).

\bigskip

\begin{figure}[t!]
    \centering
    \includegraphics[width=0.48\linewidth]{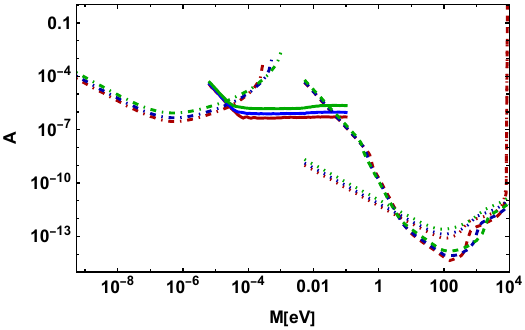}
    \includegraphics[width=0.48\linewidth]{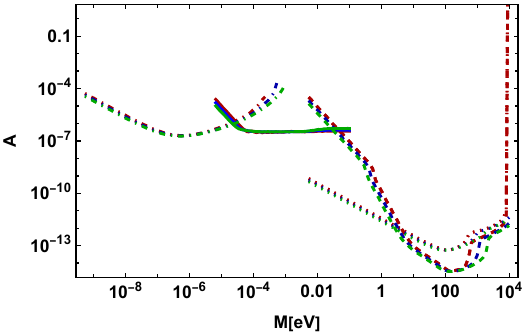}
    \caption{Median bounds on scaling parameter $A$ allowing for kinetic mixings between the DP species. We show the maximal $A$ such that the numerical median of observables is reconcilable with observation. Green: $n=40$, blue: $n=10$, red: $n=3$. Dashed: Transversal solar bounds, dotted: Longitudinal solar bounds, continuous: LSW bounds, dot-dashed: Cavendish bounds. Left: $p$=1, right: $p$=2. Sample size N=1000. $\chi_\text{max}\times\sqrt{n}=0.8$.}
    \label{fig::3}
\end{figure}

\begin{figure}[t]
    \centering
    \includegraphics[width=0.48\linewidth]{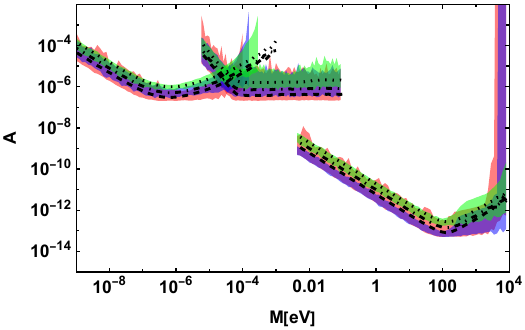}
    \includegraphics[width=0.48\linewidth]{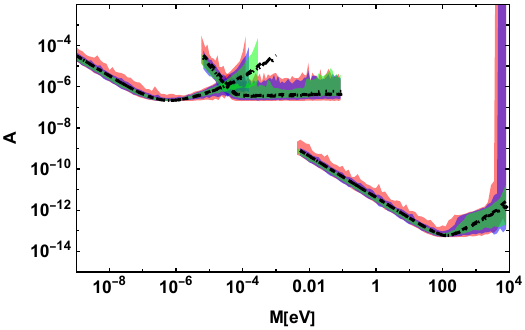}
    \caption{Bounds on the scaling parameter $A$ allowing for kinetic mixing between the DP species. From left to right the curves correspond to Cavendish and LSW experiments as well as the longitudinal photon contribution to the solar luminosity. Colored regions show all the coupling parameters $A$ for given $M$, in which at least one but not all generated parameter sets are not constrained. Also shown: Maximal value of $A$ so that the numerical median of observables is reconcilable with observation. For undiagonalized Lagrangians. Dotted/Green: $n=40$, dot-dashed/blue: $n=10$, dashed/red: $n=3$. Left: $p=1$, right: $p=2$. Sample size N=1000. $\chi_\text{max}\times\sqrt{n}=0.8$.}
    \label{extrafig::2}
\end{figure}

 The resulting ``median bounds'' from all observables and including mixing between the DP species are shown in Fig.~\ref{fig::3}. In addition, Fig.~\ref{extrafig::2}
 includes the whole range of upper bounds on $A$ resulting from at least one of the generated parameter sets. 
 
\bigskip

\begin{figure}[!t]
    \centering
    \begin{subfigure}[t]{0.48\linewidth}
        \caption*{Cavendish Voltage Ratio}
        \includegraphics[width=\linewidth]{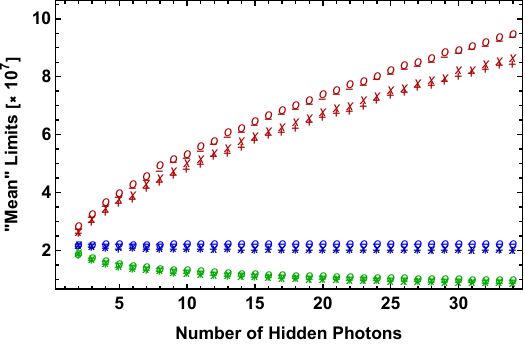}
    \end{subfigure}
    \hspace*{0.2cm}
    \begin{subfigure}[t]{0.48\linewidth}
        \caption*{LSW Detection Rate}
        \includegraphics[width=\linewidth]{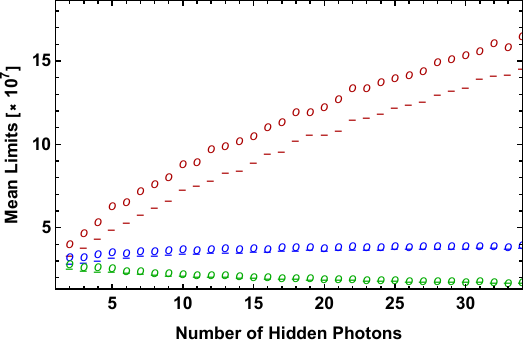}
        \vspace{5pt}
    \end{subfigure}
    \begin{subfigure}[t]{0.48\linewidth}
        \caption*{Trans. Solar Lum.}
        \includegraphics[width=\linewidth]{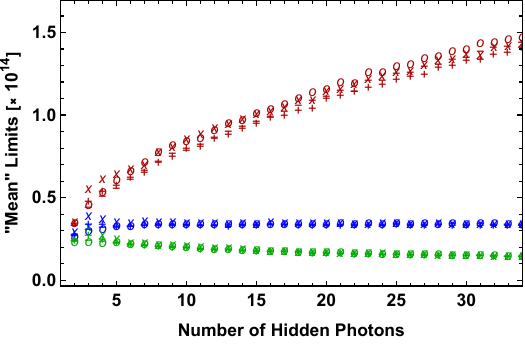}
    \end{subfigure}
    \hspace*{0.2cm}
    \begin{subfigure}[t]{0.48\linewidth}
        \centering
        \caption*{Long. Solar Lum.}
        \includegraphics[width=\linewidth]{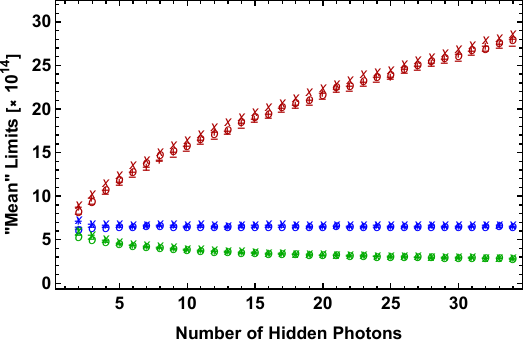}        
    \end{subfigure}
    \caption{
    $n$-dependence of the limits based on the considered observables (cf. captions). +(-): shows mean, with (without) kinetic mixing. x(o) shows median with (without) kinetic mixing. Green: $p$=4, blue: $p$=2, red: $p$=1. Scale parameter $A$ normalized to 1. Mass values around the peaks in Fig.~\ref{fig::1}, using $\chi_\text{max}\times\sqrt{n}=0.8$. Sample size N=1000. Mass scale parameters: ($M_\text{Cav}=10^{-6}$ \si{\eV}, $M_\text{LSW}=10^{-7/2}$ \si{\eV} and $M_\text{Sol}=2\times10^{2} $ \si{\eV}). }
    \label{fig::6}
\end{figure}

It is interesting to see how the limits change with rising $n$ and how this compares to the simpler case given in 
Tab.~\ref{table::1} and discussed in the previous subsection. 
For this purpose, we calculate the normalized (i.e. $A=1$) observables over many (N=1000) generated sets of parameters for a specific $M$ and different $n$ and $p$. Subsequently, the respective experimental sensitivities can be divided by the result obtained and taken to the inverse power of the power of the coupling appearing in the expressions~\eqref{eq::5}, \eqref{eq::6},\eqref{eq::8} and \eqref{eq::8b}($1/4$ for LSW, $1/2$ for all other experiments). We show the mean and median of these distributions in Fig.~\ref{fig::6} for the undiagonalized as well as the diagonalized datasets.

\begin{figure}[t!]
    \centering
    \includegraphics[scale=1.2]{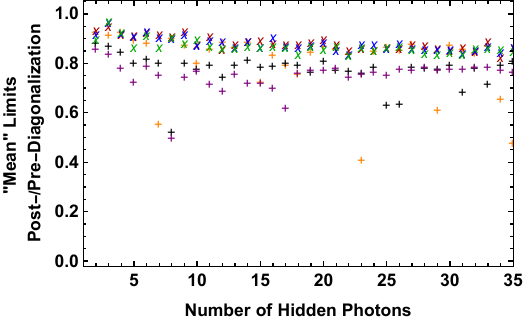}
    \caption{
    $n$-dependence of Median(x) and Mean(+) of limits on A from LSW detection rates when demanding that different choices of the couplings' powers are fixed. The quantities are normalized with their counterpart in the absence of kinetic mixing in Fig.~\ref{fig::6}. The scale parameter $A$ is normalized to 1. Mass values lie around the peaks in Fig. ~\ref{fig::1}, using $\chi_\text{max}\times\sqrt{n}=0.8$. Red: Median/$p$=1, blue: Median/$p$=2, green: Median/$p$=4, orange: Mean/$p$=1, black: Mean/$p$=2, purple: Mean/$p$=4. Sample size N=1000.}
    \label{fig::6*}
\end{figure}

We find that, regardless of the power, with respect to which we normalize, the physical observables still behave fairly similarly to the observables in the diagonalized setting. However, 
we find the two curves in Fig.~\ref{fig::6} to diverge slightly. The reason is the same as stated earlier: The diagonalization process involves a renormalization of the ``field basis vectors'' by the inverse of the kinetic matrix eigenvalues (which add to one due to the conservation of the trace). Particularly small eigenvalues can then potentially lead to disproportionately large mass or coupling parameters. The probability of this increases with the dimension of the kinetic matrix. In general, it is hard to say whether this effect increases or decreases the observable effects.  
Typically, larger coupling values amplify the observable effects in Eqs.~\eqref{eq::5}, \eqref{eq::6}, \eqref{eq::8} and \eqref{eq::8b}. Larger masses beyond the peak region lead to a reduction. 
Whether this compensates for the augmentation caused by the increase of the coupling (see, e.g., the solar lifetime curve) or not (see, e.g., the Cavendish curve) depends on the observable. For example, the expected sensitivity of LSW experiments benefits most strongly from off-diagonal kinetic couplings (while keeping diagonal couplings and masses fixed). This can be understood from the one DP photon limit which features no true peak region, and therefore larger mass values are not bound to lead to a reduction of the signal \footnote{This holds up to the point where the dark photon masses exceed the laser energy and accounting for the relatively minor reduction of the signal due to the loss of coherence between the individual contributions around $0.01~\text{eV}$.}. In this case, we also find the highest deviation between the limits based on the median and those based on the mean. As expected from Tab.~\ref{table::1} and~\ref{table::2} the $n$-dependence of the LSW detection rate is enhanced with respect to the other observables, which leads to large deviations within the considered range of dark photon numbers. We therefore show separately the $n$-behavior of these limits in the diagonal, Fig.~\ref{fig::6}, and non-diagonal case, Fig.~\ref{fig::6*}, respectively. 
The former confirms the asymptotic behavior listed in Tab.~\ref{table::1} and~\ref{table::2} for the median and mean, while the latter shows the relative change of these curves, when off-diagonal values limited by $\chi_\text{max}\times\sqrt{n}=0.8$ are switched on. In line with our previous argument we therefore find an $n$-independent improvement in the limit, see Fig.~\ref{fig::6*}.

\begin{figure}[!t]
    \centering
    \begin{subfigure}[t]{0.48\linewidth}
        \caption*{Cavendish Voltage Ratio}
        \includegraphics[width=\linewidth]{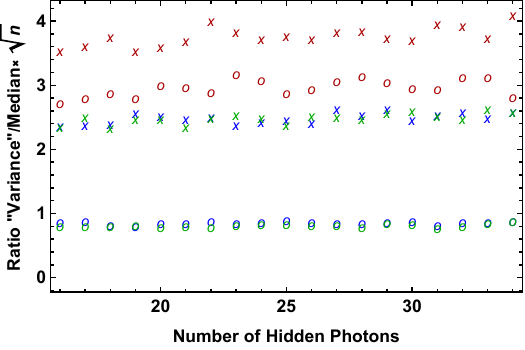}
    \end{subfigure}
    \begin{subfigure}[t]{0.48\linewidth}
        \caption*{LSW Detection Rate}
        \includegraphics[width=\linewidth]{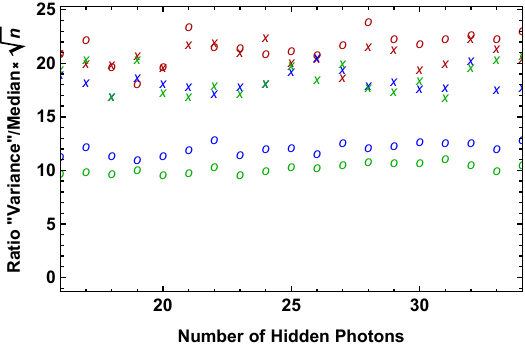}
        \vspace{5pt}
    \end{subfigure}
    \begin{subfigure}[t]{0.48\linewidth}
        \caption*{Trans. Solar Lum.}
        \includegraphics[width=\linewidth]{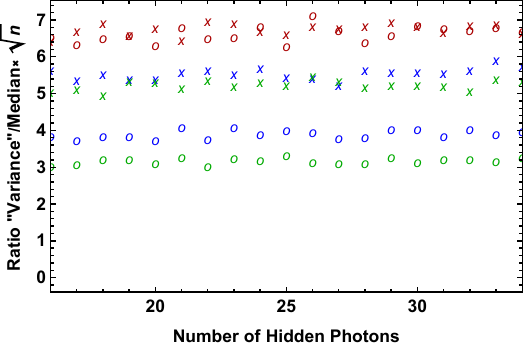}
    \end{subfigure}
    \begin{subfigure}[t]{0.48\linewidth}
        \centering
        \caption*{Long. Solar Lum.}
        \includegraphics[width=\linewidth]{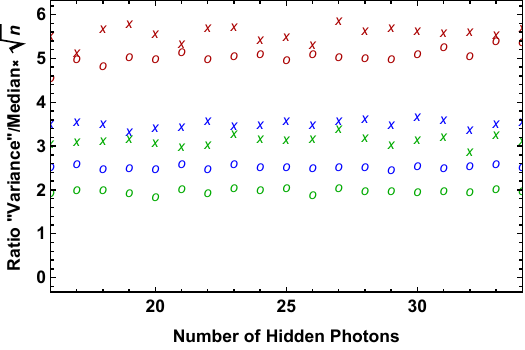}
    \end{subfigure}
    \caption{For the constraints under consideration: $\frac{\text{“Standard Deviation"}}{\text{Median}}\times\sqrt{n}$ of the observables ~\eqref{eq::5}, \eqref{eq::6} and \eqref{eq::8} as a function of DP number $n$. “Standard Deviation" is defined as the spread of the 90\% confidence interval. Sample size N=1000. Using $\chi_\text{max}\times\sqrt{n}=0.8$. Green: $p$=4, blue: $p$=2, red: $p$=1. x(o) indicates the inclusion (omission) of kinetic mixing. Mass scale parameters ($M_\text{Cav}=10^{-6}$ \si{\eV}, $M_\text{LSW}=10^{-7/2}$ \si{\eV} and $M_\text{Sol}=2\times10^{2} $ \si{\eV}).}
    \label{extrafig::5}
\end{figure}

As we can see from Fig.~\ref{extrafig::5}, a suitable median-adapted analogue to $\frac{\textbf{variance}}{\textbf{mean}^2}$ introduced in Eq.~\eqref{eq:variancemean} also is $\propto\frac{1}{n}$.

Obviously, this only tells us how the average (with respect to a power p) of these parameters changes after diagonalization and the final effect on the observable is more intricate since it also depends on how the $m_i^2$ are distributed on the different $\epsilon_i$ (the way they transform is not identical). Nonetheless, this still provides a rough idea of the different behaviors with and without DP mixing as shown in Fig.~\ref{fig::6}

\begin{figure}[!t]
    \centering
    \begin{subfigure}[t]{0.48\linewidth}
        \caption*{Cavendish,  $p$=1}
        \includegraphics[width=\linewidth]{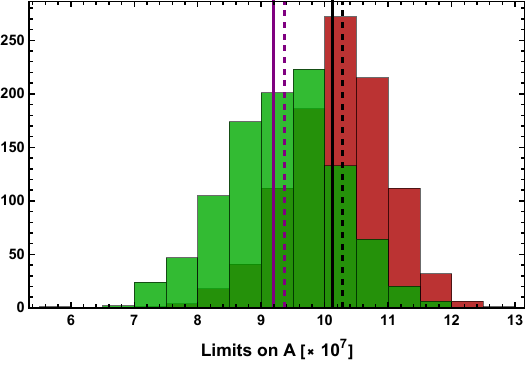}
    \end{subfigure}
    \begin{subfigure}[t]{0.48\linewidth}
        \caption*{LSW, $p$=4}
        \includegraphics[width=\linewidth]{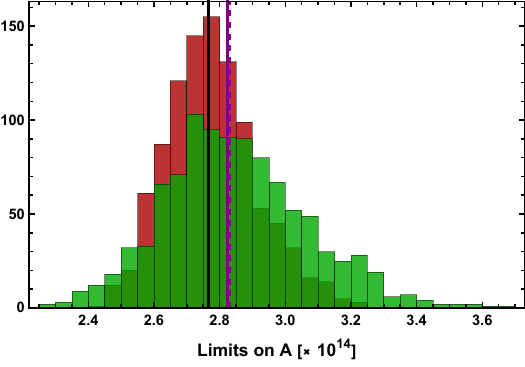}
    \end{subfigure}
    \\
    \vspace{.5cm}
    \begin{subfigure}[t]{0.48\linewidth}
        \caption*{Solar T,  $p$=2}
        \includegraphics[width=\linewidth]{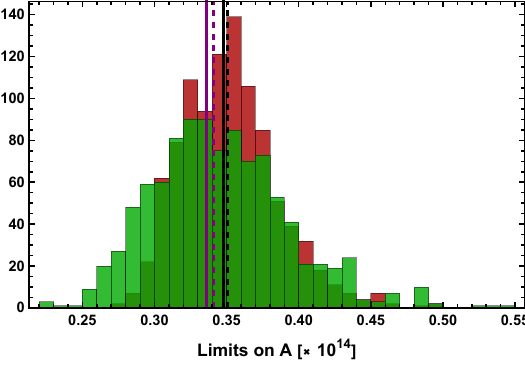}
    \end{subfigure}
    \begin{subfigure}[t]{0.48\linewidth}
        \centering
        \caption*{Solar L,  $p$=2}
        \includegraphics[width=\linewidth]{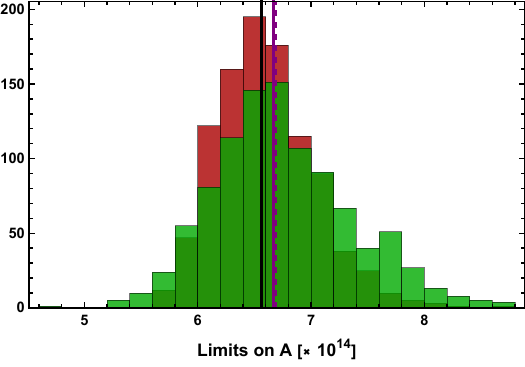}
    \end{subfigure}
 \caption{Examples of the statistic distributions of the physical effects in different settings (cf. captions) when generating diagonalized (red) and undiagonalized couplings (green) pertaining to certain $M$ ($M_\text{Cav}=10^{-6}$ \si{\eV}, $M_\text{LSW}=10^{-7/2}$ \si{\eV} and $M_\text{Sol}=2\times 10^{2} $ \si{\eV}) and $A$ normalized to 1. Dashed (connected) lines give the median (mean) of the distributions. Purple: DP mixing, black: No DP mixing. Sampling size N=1000. Histograms show number of samples per bin. }
    \label{fig::7}
\end{figure}

Finally, let us compare the distributions of the random variables Eqs.~\eqref{eq::5}, \eqref{eq::6} and \eqref{eq::8} (i.e., the physical effects) when generating the already diagonalized couplings and when generating the “bare” undiagonalized couplings. Since the diagonalization includes a renormalization by the inverse of the eigenvalues of the kinetic matrix in \eqref{eq::2}, we might divide by values close to 0 for large enough $n$. In particular, we expect the coupling to be larger post-diagonalization than pre-diagonalization, which leads to larger means and variances. Some exemplary distributions for each of the considered physical effects are shown in Fig. \ref{fig::7}. For each type of observable, we chose $M$ to lie within the most constrained regions seen in Fig.~\ref{fig::3} ($M_\text{Cav}=10^{-6}$ \si{\eV}, $M_\text{LSW}=10^{-7/2}$ \si{\eV} and $M_\text{Sol}=2\times 10^{2} $ \si{\eV}), $n$ is 40 and the maximum value of the $\chi$ is given by 0.25. We find the outcome to agree with the expectations we argued for above. Notably the distributions are more spread out when allowing for mixing between the DP species as this generates additional statistical variations.\footnote{Note that the mean value shown in Fig.~\ref{fig::7} is calculated with the sample distribution and may differ from the one obtained for infinite sample size, due to outliers.}

\section{Brief conclusions}\label{sec::disc}
In this work, we have investigated the phenomenology of a scenario with multiple dark photons, kinetically mixed with the Standard Model photons as well as among themselves. 
For the values of the masses and mixing parameters, we have chosen a simple anarchical approach: The parameters are random variables following a distribution function that contains a single parameter determining the overall scale of the random variables.

Comparing with experimental and observational results, we have focused on three exemplary types of  constraints on dark photons:
Cavendish-type experiments probing the Coulomb form of the electromagnetic potential, light-shining-through-walls (LSW) experiments and observations of solar energy loss. We have shown how to extend these constraints from the case of a single dark photon.
While there exist other constraints, this selection already neatly demonstrates crucial features of the generalization to the case with multiple dark photons.

 The signals in Cavendish and solar energy loss are  additive, i.e., when adding an extra DP\footnote{Strictly speaking, in this statement we assume that the extra dark photon state mixes only with the ordinary photon and not with other dark photon states. Such a mixing could, in principle, reduce the mixing of the other dark photons with the ordinary photon. Our numerical results nevertheless show that this is usually not the case. } to an already existing set, the constraint gets stronger. In contrast, this is not necessarily the case for the limits imposed by LSW experiments. There, destructive interference in the signal is possible.
Nevertheless, statistically, this is unlikely to happen.

The resulting limits on the dark photons generally exhibit a similar shape as in the single photon case, featuring a relatively mild $n$ dependence. Analytical results obtained for the case without mixing can be found in Tabs.~\ref{table::1} and~\ref{table::2}.
It is noteworthy that the degree of similarity to the single dark photon case hinges on whether the limits are based on the median or the mean of the distributions. For example, in the case of the solar energy loss, and looking at the region of large masses, the latter is affected by rare but possible unusually light dark photons that can still be produced at solar temperatures whereas the more typical heavier ones are strongly Boltzmann suppressed.
For this reason, if we aim to give a ``single-line'' limit, the median seems more suitable than the mean.

When allowing for kinetic mixings between the different dark photon species a new feature arises. Keeping the overall scale of the mixings between the different DPs fixed, it becomes increasingly likely that the kinetic matrix has one or more negative eigenvalues, rendering the theory problematic.  Using Wigner's semicircle law we argue that this effect scales with $\sqrt{n}$ and normalize our mixing parameter accordingly.
Choosing moderately large, but still stable values for the so normalized mixings we find a quantitative impact on the obtained limits, but the qualitative features remain similar.

\begin{appendix}
\section{Diagonalization of DP-field space}\label{AppendixA}
In this section, we carry out the diagonalization in \ref{sec::2.1} more explicitly. We start with \eqref{eq::2} and, as mentioned previously, multiply the entire equation by the matrix 
\begin{gather}\label{eq::10}                
    \mathcal{S_0}=\begin{pmatrix}
        1 & -\epsilon_1 & -\epsilon_2 & ... \\
       0& 1 &0 &\\
        0 & 0 & 1 \\
        ...\\
    \end{pmatrix}
\end{gather}
from the right and its transposed from the left. This directly leads to \eqref{eq::3}. Then, we remark that the $n\times n$ ``kinetic'' submatrix is symmetric, so it can be diagonalized by a basis change related to the orthogonal matrix $\mathcal{T}_1$. The new and old field bases are thus related by the matrix,
\begin{gather}\label{eq::10+1}                
    \mathcal{S_1}=
    \begin{pmatrix}
        1 & 0 \\
       0& \mathcal{T}_1    
    \end{pmatrix}\,.
\end{gather}

Upon a multiplication of the entire equation by the transpose of $\mathcal{S}_1$ from the left, the first matrix in \eqref{eq::2} is then fully diagonalized. We wish to transform it to the identity matrix by renormalizing all fields by their corresponding eigenvalues $v_i$, leading to a field transformation matrix.

\begin{gather}\label{eq::11}                
    \mathcal{S}_2=\textbf{diag}\begin{pmatrix}
        v_0 &v_1 &... & v_{n+1}    
    \end{pmatrix}
    \equiv
    \begin{pmatrix}
        1 &0\\
        0& \mathcal{T}_2
    \end{pmatrix}\,.
\end{gather}
As always, the entire equation is multiplied from the left by $\mathcal{S}_2^T$. At this point, since the $\mathcal{S}_i$ are all either orthogonal or diagonal, the mass matrix and the matrix that multiplies $\pi_a$ are still symmetric and the mass matrix can be diagonalized by yet another basis change represented by the matrix $\mathcal{T}_3$, which leads to,
\begin{gather}\label{eq::12}                
    \mathcal{S}_3=
    \begin{pmatrix}
        1 & 0 \\
       0& \mathcal{T}_3    
    \end{pmatrix}\,.
\end{gather}
Applying the by now well established transformation of multiplying by $\mathcal{S}_3^T$ from the left, we obtain,
\begin{gather}\label{eq::13}                
    \left[
    \Box
    \begin{pmatrix}
        1 & 0 & 0 & ... \\
        0 & 1 & 0 &\\
        0 & 0 & 1 \\
        ...\\
    \end{pmatrix}
    + 
    \pi_a
    \begin{pmatrix}
    1 & -\Vec{\epsilon^*}^T\\
    -\Vec{\epsilon^*} & \Vec{\epsilon^*}\Vec{\epsilon^*}^T
    \end{pmatrix}
    +
    \begin{pmatrix}
    0 & 0 & 0 & ...\\
    0 & {m_1^2}^* & 0 \\
    0 & 0& {m_2^2}^* \\
    ...\\
    \end{pmatrix}
    \right]
    \begin{pmatrix}
        A'' \\ X_1'' \\ X_2''\\...
    \end{pmatrix}
    =
    j
    \begin{pmatrix}
        1 \\ -\vec{\epsilon*}
    \end{pmatrix}\,.
\end{gather}
Here, $A'', X_i''$ are the fields after having carried out the final change of basis in \eqref{eq::12}. Performing the inverse transformation (to $\mathcal{O}(\epsilon)$) pertaining to \eqref{eq::10} with $\epsilon_i$ exchanged by $\epsilon_i^*$ removes the last $n$ current entries and finally returns the result \eqref{eq::4}. It is quickly verified that,
\begin{gather}\label{eq::14}                
    \vec{\epsilon}^*=\mathcal{T}_3\mathcal{T}_2\mathcal{T}_1\vec{\epsilon}\,.
\end{gather}

Finally, note that we could have alternatively applied the basis change
\begin{gather}\label{eq::10*}                
    \mathcal{S_0}=\begin{pmatrix}
        1 & 0 & 0 & ... \\
       \epsilon_1& 1 &0 &\\
        \epsilon_2 & 0 & 1 \\
        ...\\
    \end{pmatrix}.
\end{gather}
to~\eqref{eq::2}.
This also succeeds at removing the kinetic mixing between $A^\mu$ and the DP fields but electromagnetic currents will not be charged under them. In exchange, there is now a mixing in the mass matrix. This parametrization is particularly useful in an opaque medium, as the wall in LSW experiments, c.f. Sec.~\ref{sec:2.3}. In this setting, $A^\mu$ itself is already massive and its mass can be assumed to be parametrically larger than the DP masses. As a consequence, the mass matrix is strongly hierarchical and the mixing angles between $A^\mu$ and the individual $X^\mu$ is suppressed, such that the equation of motion may be approximated as,  
\begin{gather}\label{eq:app.EOMlargeAmass}
    \left[
    \Box
    \begin{pmatrix}
        1 & 0 & 0 & ... \\
        0 & 1 & 0 &\\
        0 & 0 & 1 \\
        ...\\
    \end{pmatrix}
    +
    \begin{pmatrix}
    \pi_a & 0 & 0& ...\\
    0& {m_1^2}^* & 0 \\
    0 & 0& {m_2^2}^* \\
    ...\\
    \end{pmatrix}
    \right]
    \begin{pmatrix}
        A'' \\ X_1'' \\ X_2''\\...
    \end{pmatrix}
    =
    j
    \begin{pmatrix}
        1 \\ 0 \\ ...
    \end{pmatrix}\,,
\end{gather}
to first order in $\epsilon^*$ and $\sqrt{\pi_a/{m_i^*}^2}$.

\section{N-Behavior of means and variances of the observables}\label{AppendixB}

Here we show, as promised in Sec.~\ref{sec::3.1}, the leading $n$-behaviour of the different observables' means and variances used to obtain the constraints. The relation between the results Tab. \ref{table::1} and Tab. \ref{table::2} to the expressions given here for the mean of the physical effect is then described by~\eqref{eq::boundscaling}.
Explicitly the experimental thresholds are $\gamma_\text{Cav}$=8.0\si{\times 10^{-15}}, $\gamma_\text{LSW}$=2.6\si{\times 10^{-26}} and $\gamma_\text{Solar}$=0.1 given in~\cite{boundary3,LSWAlps,redondo08,Serenelli}, since we normalized the random variables associated to solar energy loss arguments in the tables below by $L_{Sol}=$1.6\si{\times 10^{30}\eV} \cite{Serenelli} and it is commonly postulated that the DP channel should not account for more than 10\% of the solar energy loss. As in the main text, we distinguish the coherent ($\Delta m^2 \ll \omega/L_\text{w}$) and incoherent ($\Delta m^2 \gg \omega/L_\text{w}$) regimes in the context of LSW experiments.

\begingroup
\setlength{\tabcolsep}{10pt} 
\renewcommand{\arraystretch}{1.5} 
\begin{table}[t] 
\begin{adjustbox}{width=\textwidth, center}
    \centering
  \begin{tabular}{c|ccc}    
    &$p$=1 & $p$=2 & $p$=4 \\
    \hline
    Cavendish  &$n^{-1}$ 1.5\si{\times 10^{13}} $M^2$/\si{\eV^2} & 7.5\si{\times 10^{12}} $M^2$/\si{\eV^2} & $\sqrt{n}$ 6.6\si{\times 10^{12}}$M^2$/\si{\eV^2} \\
    LSW &$n^{-2}$ 1.0\si{\times 10^{29}} $M^8$/\si{\eV^8} & 2.6\si{\times 10^{28}} $M^8$/\si{\eV^8}& $n$ 2.0\si{\times 10^{28}} $M^8$/\si{\eV^8}  \\
   Sun (T)  & $n^{-1}$ 8.4\si{\times 10^{17}} $M^4$ /\si{\eV^4} & 4.2\si{\times 10^{17}} $M^4$ /\si{\eV^4} & $\sqrt{n}$ 3.7\si{\times 10^{17}} $M^4$ /\si{\eV^4} \\ 
    Sun (L) & $n^{-1}$ 1.1\si{\times 10^{22}} $M^2$ /\si{\eV^4}& 5.7\si{\times 10^{21}} $M^2$ /\si{\eV^4}&$\sqrt{n}$ 5.1\si{\times 10^{21}} $M^4$ /\si{\eV^4}\\ 
  \end{tabular}
\end{adjustbox}
  \caption{Proportionality between DP number $n$ and mean of the different physical effects for probability distribution (\ref{eq::9}), given arbitrary $A$ and $M$, small mass limit\label{table::B1}, resulting constraints scale inversely with given expressions as described by~\eqref{eq::boundscaling}}
\end{table}
\endgroup

\begingroup
\setlength{\tabcolsep}{10pt} 
\renewcommand{\arraystretch}{1.5} 
\begin{table}[t!] 
\begin{adjustbox}{width=\textwidth, center}
\centering
  \begin{tabular}{c|ccc}    
    &$p$=1 & $p$=2 & $p$=4 \\
    \hline
    
    Cavendish &$n^{-3}$ 1.3\si{\times 10^{27}} $M^4$/\si{\eV^4} &$n^{-1}$ 5.6\si{\times 10^{25}} $M^4$/\si{\eV^4}  & 1.3 \si{\times 10^{25}} $M^4$/\si{\eV^4} \\
    
    LSW & $n^{-5}$ 1.1\si{\times 10^{60}}  $\left(M/\text{\si{\eV}}\right)^{16}$  & $n^{-1}$ 1.6\si{\times 10^{58}}  $\left(M/\text{\si{\eV}}\right)^{16}$ & $n$ 3.9\si{\times 10^{57}} $\left(M/\text{\si{\eV}}\right)^{16}$\\ 
    Sun (T)  &$n^{-3}$ 2.1\si{\times 10^{37}} $M^8$/\si{\eV^8} &$n^{-1}$ 1.6 \si{\times 10^{36}} $M^8$/\si{\eV^8} & 7.4\si{\times 10^{35}} $M^8$/\si{\eV^8}\\  
    Sun (L)& $n^{-3}$ 7.8\si{\times 10^{44}} $M^2$/\si{\eV^2}& $n^{-1}$ 3.2\si{\times 10^{43}} $M^2$/\si{\eV^2}& 7.6\si{\times 10^{42}} $M^2$/\si{\eV^2}\\ 
  \end{tabular}
\end{adjustbox}
\caption{Proportionality between DP number $n$ and variance of the different physical effects for probability distribution (\ref{eq::9}), given arbitrary $A$ and $M$, small mass limit resulting constraints scale inversely with given expressions as described by~\eqref{eq::boundscaling}. \label{table::B2}}
\end{table}
\endgroup
\begingroup
\setlength{\tabcolsep}{10pt} 
\renewcommand{\arraystretch}{1.5} 
\begin{table}[t!] 
    \begin{adjustbox}{width=\textwidth, center}
    \centering
  \begin{tabular}{c|ccc}    
    &$p$=1 & $p$=2 & $p$=4 \\
    \hline
    
    Cavendish &$n^{-1}$ 2.3\si{\times 10^{-11}\eV^2\per}$M^2$ & 1.2\si{\times 10^{-11}\eV^2\per}$M^2$ & $\sqrt{n}$ 1.0 \si{\times 10^{-11}\eV^2\per}$M^2$ \\
    
    LSW ($\Delta m^2 \ll \omega/L_\text{w}$) & 4 $n^{-2}$ & 1 &0.8 $n$ \\

    LSW ($\Delta m^2 \gg \omega/L_\text{w}$) &$144~ n^{-3}$  & $12~n^{-1}$  & 6 \\
    
    Sun (T)  &$n^{-1}$ 2.2\si{\times 10^{33} \eV^2\per}$M^2$ & 1.1 \si{\times 10^{33} \eV^2\per}$M^2$ & $\sqrt{n}$ 9.9 \si{\times 10^{32} \eV^2\per}$M^2$ \\

    Sun (L) & $n^{-1}$ 4.9\si{\times 10^{30} \eV^2\per}$M^2$& 2.4\si{\times 10^{30} \eV^2\per}$M^2$ & $\sqrt{n}$ 2.2\si{\times 10^{30} \eV^2\per}$M^2$ \\  
  \end{tabular}
    \end{adjustbox}
  \caption{Proportionality between DP number $n$ and mean of the different physical effects for probability distribution (\ref{eq::9}), given arbitrary $A$ and $M$, large mass limit. \label{table::B3}}
\end{table}
\endgroup
\begingroup
\setlength{\tabcolsep}{10pt} 
\renewcommand{\arraystretch}{1.5} 
\begin{table}[t!]
    \begin{adjustbox}{width=\textwidth, center}
    \centering
  \begin{tabular}{c|ccc}    
    &$p$=1 & $p$=2 & $p$=4 \\
    \hline
    Cavendish &$n^{-3}$ 4.8\si{\times 10^{-12}\eV^2\per M^2} &$n^{-1}$ 4.0\si{\times 10^{-13}\eV^2\per M^2} & 2.0 \si{\times 10^{-13}\eV^2\per M^2}\\
    
    LSW ($\Delta m^2 \ll \omega/L_\text{w}$)&1024.0 $n^{-5}$  &20.0 $n^{-1}$ & 7.9~$n$\\

    LSW ($\Delta m^2 \gg \omega/L_\text{w}$) &$20736.0~ n^{-6}$  & $144.0~n^{-2}$  & 36.0 \\
    
    Sun (T)  &$n^{-3}$ 4.3 \si{\times 10^{62} \eV^2\per M^2} &$n^{-1}$ 3.6 \si{\times 10^{61} \eV^2\per M^2} & 1.8 \si{\times 10^{61} \eV^2\per M^2}\\  
    Sun (L) & $n^{-3}$ 1.4\si{\times 10^{57} \eV^2\per M^2}& $n^{-1}$ 1.2\si{\times 10^{56} \eV^2\per M^2}& 5.8\si{\times 10^{55} \eV^2\per M^2}\\ 
  \end{tabular}
  \end{adjustbox}
  \caption{Proportionality between DP number $n$ and variance of the different physical effects for probability distribution (\ref{eq::9}), given arbitrary $A$ and $M$, large mass limit. \label{table::B4}}
\end{table}
\endgroup
\end{appendix}

\pagebreak
\bibliographystyle{utphys}
\bibliography{bib}
\end{document}